\documentclass[conference]{IEEEtran}
\pdfoutput=1  
\IEEEsettextheight{0.8in}{1.0in}
\IEEEsettopmargin{t}{0.8in}
\IEEEoverridecommandlockouts

\usepackage[font=it,labelfont=bf]{caption}
\usepackage{cite}
\usepackage{amsmath,amssymb,amsfonts}
\usepackage{algorithmic}
\usepackage{graphicx}
\usepackage{textcomp}
\usepackage{array}
\usepackage{subcaption}
\usepackage{xcolor}
\usepackage{xurl}
\usepackage{multirow}
\usepackage{siunitx}

\def\BibTeX{{\rm B\kern-.05em{\sc i\kern-.025em b}\kern-.08em
    T\kern-.1667em\lower.7ex\hbox{E}\kern-.125emX}}

\definecolor{warningred}{RGB}{220, 53, 69}       
\definecolor{warningorange}{RGB}{0, 123, 255}    
\definecolor{warningpurple}{RGB}{138, 43, 226}    
\definecolor{warninggreen}{RGB}{70, 161, 4}       
\definecolor{warningteal}{RGB}{0, 180, 180}       

\newif\ifcomment

\commenttrue 

\ifcomment
\newcommand{\commentmcv}[1]{{\emph{\textcolor{warningred}{[MCV: #1]}}}}
\newcommand{\commentsb}[1]{{\emph{\textcolor{warningorange}{[SB: #1]}}}}
\newcommand{\commentca}[1]{{\emph{\textcolor{warningteal}{[CA: #1]}}}}
\newcommand{\commentab}[1]{{\emph{\textcolor{warningpurple}{[Avhi: #1]}}}}
\newcommand{\commentap}[1]{{\emph{\textcolor{warninggreen}{[Apala: #1]}}}}

\else
\newcommand{\commentmcv}[1]{}
\newcommand{\commentsb}[1]{}
\newcommand{\commentca}[1]{}
\newcommand{\commentab}[1]{}
\newcommand{\commentap}[1]{}
\fi

\begin{document}

\title{Waves on the Walls:\\ Empirical Characterization of mmWave Lateral Waves for Enhanced Indoor Coverage}

\author{
\IEEEauthorblockN{Apala Pramanik$^{1}$,
  Avhishek Biswas$^{1}$,
  Sasitharan Balasubramaniam$^{1}$,
  Christos Argyropoulos$^{2}$,
  Mehmet C. Vuran$^{1}$}
\IEEEauthorblockA{$^1$\textit{School of Computing,
University of Nebraska--Lincoln}, Lincoln, NE, USA\\
\{apramanik2, abiswas3\}@huskers.unl.edu, \{sasi, mcv\}@unl.edu}
\IEEEauthorblockA{$^2$\textit{Electrical Engineering and Computer Science,
The Pennsylvania State University}, University Park, PA, USA\\
cfa5361@psu.edu}
}
\maketitle
\bstctlcite{IEEEexample:BSTcontrol}
\begin{abstract}

High-frequency millimeter-wave (mmWave) communication systems are constrained by the surrounding environment, where walls are traditionally treated as obstacles that block or reflect signals indoors. Consequently, current beamforming strategies are tailored to circumvent these obstructions. In this paper, a paradigm shift is introduced that leverages lateral wave propagation along building interfaces to extend mmWave coverage. Unlike traditional reflections, lateral waves travel along the boundary between two media of different refractive indices and decay algebraically with distance, offering a potential alternative path for mmWave connectivity. While well-established at low frequencies in natural media, the existence of lateral waves at mmWave frequencies along engineered building materials has not been demonstrated before. To this end, the first experimental characterization of mmWave lateral waves along a wall is reported. Extensive controlled measurements are employed to characterize the signal-grazing geometry and to establish a frequency and distance-dependent path-loss model for this phenomenon. The results provide the first empirical foundation for a new class of interface-guided mmWave links.

\end{abstract}

\vspace{-0.2cm}
\section{Introduction}
\label{sec:introduction}
The rapid growth in demand for high-capacity broadband services has accelerated the adoption of millimeter-wave (mmWave) communications as a key technology for next-generation wireless networks. Among its emerging applications, Fixed Wireless Access (FWA) delivers last-mile broadband over a wireless link between a fixed base station and a stationary customer-premises terminal. It serves users inside a home or office, in place of running fiber or copper to each building. 
By leveraging existing cellular infrastructure and eliminating the need to deploy fiber or copper to individual buildings, FWA has emerged as one of the fastest-growing services of the 5G era~\cite{oproiu2018fwa}. The terminal it serves is positioned indoors, so an FWA deployment is as much an indoor radio system as an access link, and the same infrastructure that pushes broadband into a building is well-placed to sense activity within it.
The capacity demands of FWA are increasingly met using mmWave bands. Multiple gigahertz of contiguous spectrum above 24\,GHz offer data rates unattainable at sub-6\,GHz frequencies~\cite{rappaport2013mmwave}. This makes mmWave communications particularly well suited to FWA, where stationary terminals often maintain fixed, line-of-sight links to the base station, which suits the narrow, high-gain directional beams that mmWave links require.
\begin{figure}[t!]
    \centering
    \includegraphics[width=0.9\columnwidth]{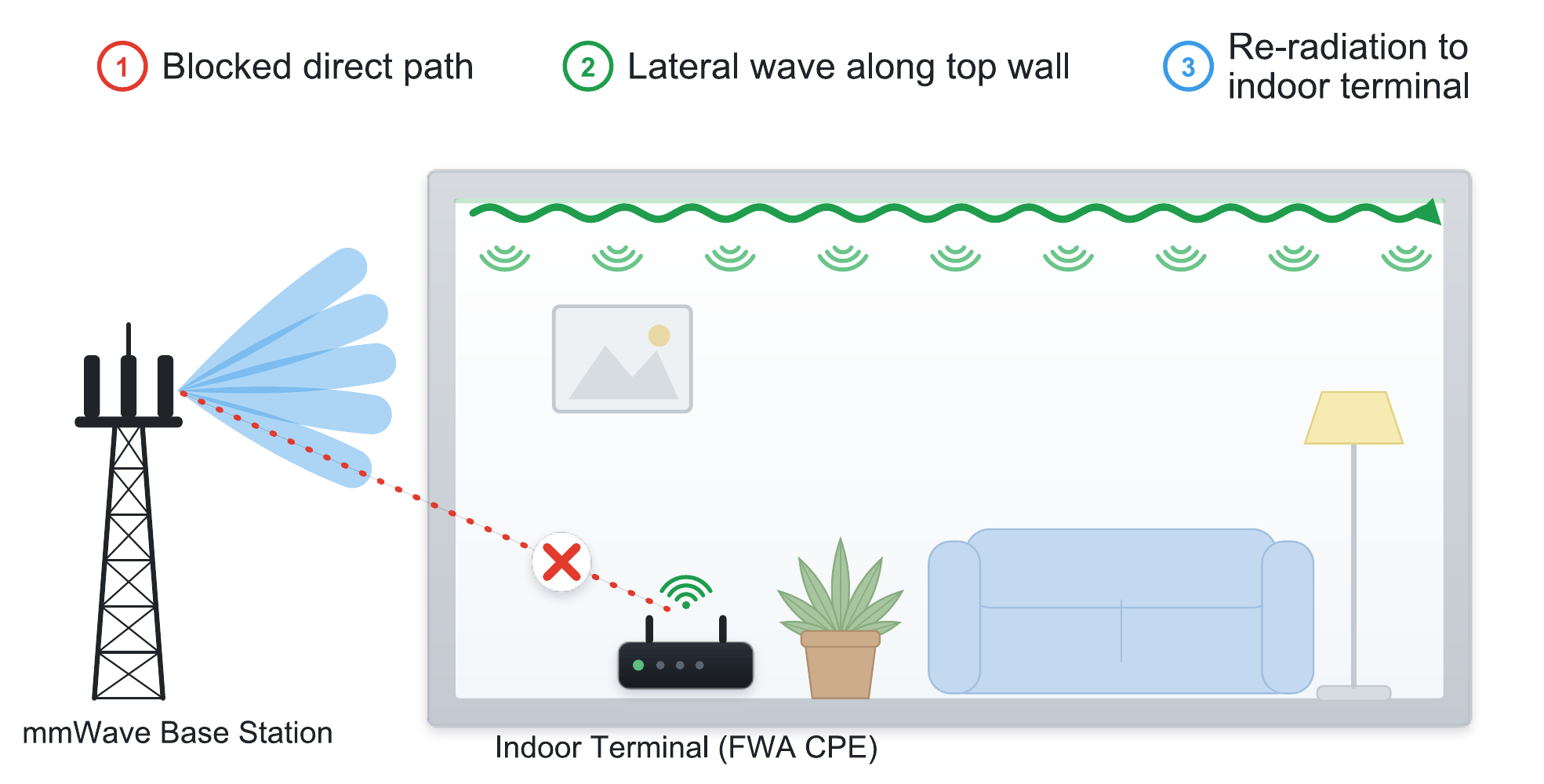}
    \caption{Lateral wave propagation in an indoor scene.}
    \label{fig:overview}
    \vspace{-2em}
\end{figure}
Despite these advantages, extending mmWave FWA coverage to indoor users (Fig.~\ref{fig:overview}) remains a significant challenge. The walls and partitions that define an indoor space attenuate mmWave signals by tens of decibels~\cite{hosseini2021attenuation,du2021sub}, making terminals behind them difficult to reach. Furthermore, mmWave links are highly directional and susceptible to blockage, where even a passing human body can drop a 60\,GHz link by more than 30\,dB~\cite{slezak2018blockage}. Moreover, maintaining a link requires tightly aligned transmit and receive beams, whose beam training adds substantially to the latency and control overhead~\cite{biswas2026secon}.

Current research approaches typically treat walls as components of the wireless system, either by engineering their reflective properties or by analyzing their scattering. For example, reconfigurable intelligent surfaces (RIS) attempt to mitigate indoor blockage by mounting tunable, engineered arrays on walls to redirect signals~\cite{di2020smart,wu2020ris}. Alternatively, the wall may also serve as a sensing medium, where analyzing strong reflection and scattering of mmWave signals can reveal the presence, motion, and activity of occupants within a room~\cite{zhang2023mmsensing}. 


In this paper, we focus on an alternative phenomenon that exploits \textit{lateral waves} traveling along wall surfaces to extend mmWave connectivity. Fundamentally, lateral waves arise at any planar boundary between media with different refractive indices, such as an air-wall interface.
Lateral waves, the electromagnetic analog of seismic head waves~\cite{cerveny1971theory}, are generated when an electromagnetic source illuminates an interface from a higher-refractive-index medium at or near the critical angle of total internal reflection~\cite{clough1976electromagnetic}. The resulting energy refracts to propagate along the interface (e.g., wall) within the lower-loss medium (e.g., air) and continuously re-radiates back into the denser medium~\cite{sommerfeld1909ausbreitung,banos1966dipole,king2012lateral}. Importantly, since most of the wave path resides in the low-loss medium, the signal decays only algebraically with distance, avoiding the exponential absorption of a direct path. This property has been extensively exploited for wireless sensor networks underground~\cite{dong2011channel,salam2019theoretical}, through forests~\cite{tamir1967radio},  through glacial ice~\cite{evans1963radio,clough1976electromagnetic}, and underwater~\cite{smolyaninov2018surface}.
While this phenomenon is well-established at low frequencies in natural media~\cite{tamir1967radio,dong2011channel,evans1963radio,smolyaninov2018surface}, its viability at mmWave frequencies along engineered building materials, whose layer thickness, surface roughness, and material dispersion are comparable to the wavelength, remains an open question. Current studies of mmWave propagation through building materials focus on penetration loss and specular/diffuse reflection~\cite{hosseini2021attenuation,jansen2008impact,jansen2011diffuse,han2022terahertz}, leaving lateral-wave propagation paths uncharacterized. 

In this paper, we report, to the best of our knowledge, the first experimental demonstration of lateral waves at mmWave frequencies along a building-material interface. We develop a path-loss model that captures the scaling laws of this wave with respect to both frequency and distance. The model is validated through rigorous controlled experimental measurements. Our results provide the empirical foundation for a new class of interface-guided mmWave networks that leverage wall surfaces as waveguides rather than mere obstacles. 

The remainder of this paper is organized as follows. We review related work in Section~\ref{sec:related_works}. In Section~\ref{sec:overview}, we experimentally demonstrate the lateral waves at mmWave frequencies. Accordingly, we develop a frequency and distance-dependent path-loss model for mmWave lateral waves in Section~\ref{sec:channel_model}, and validate it against the measured data in Section~\ref{sec:evaluation}. The paper is concluded in Section~\ref{sec:conclusion}.
\vspace{-0.10cm}
\section{Related Work}
\label{sec:related_works}

\begin{figure*}[ht!]
  \centering
  \begin{subfigure}[t]{0.29\linewidth}
    \centering
    \includegraphics[width=\linewidth]{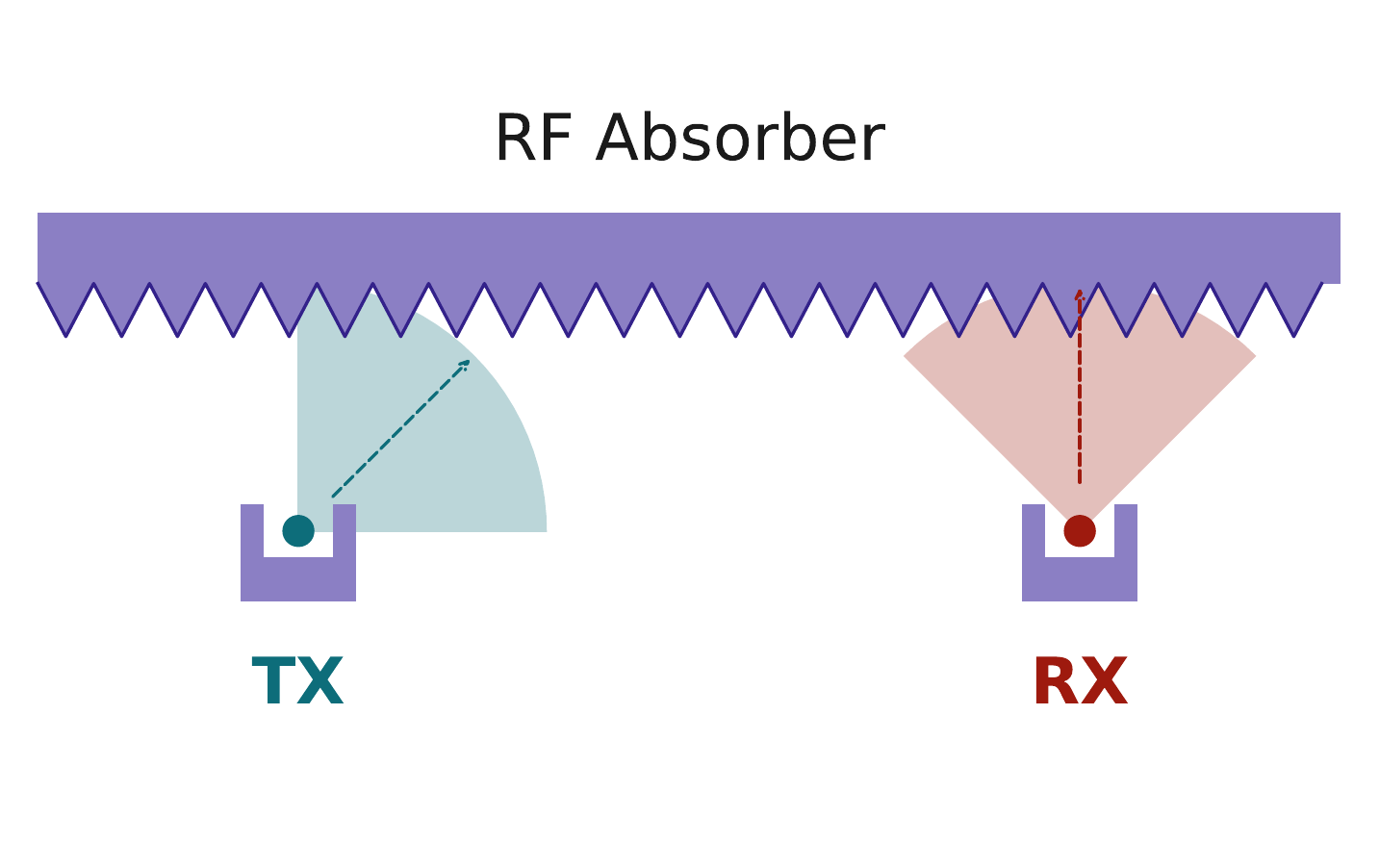}
    \caption{RF absorber only (AO).}
    \label{fig:cfg_ao}
  \end{subfigure}
  \hfill
  \begin{subfigure}[t]{0.29\linewidth}
    \centering
    \includegraphics[width=\linewidth]{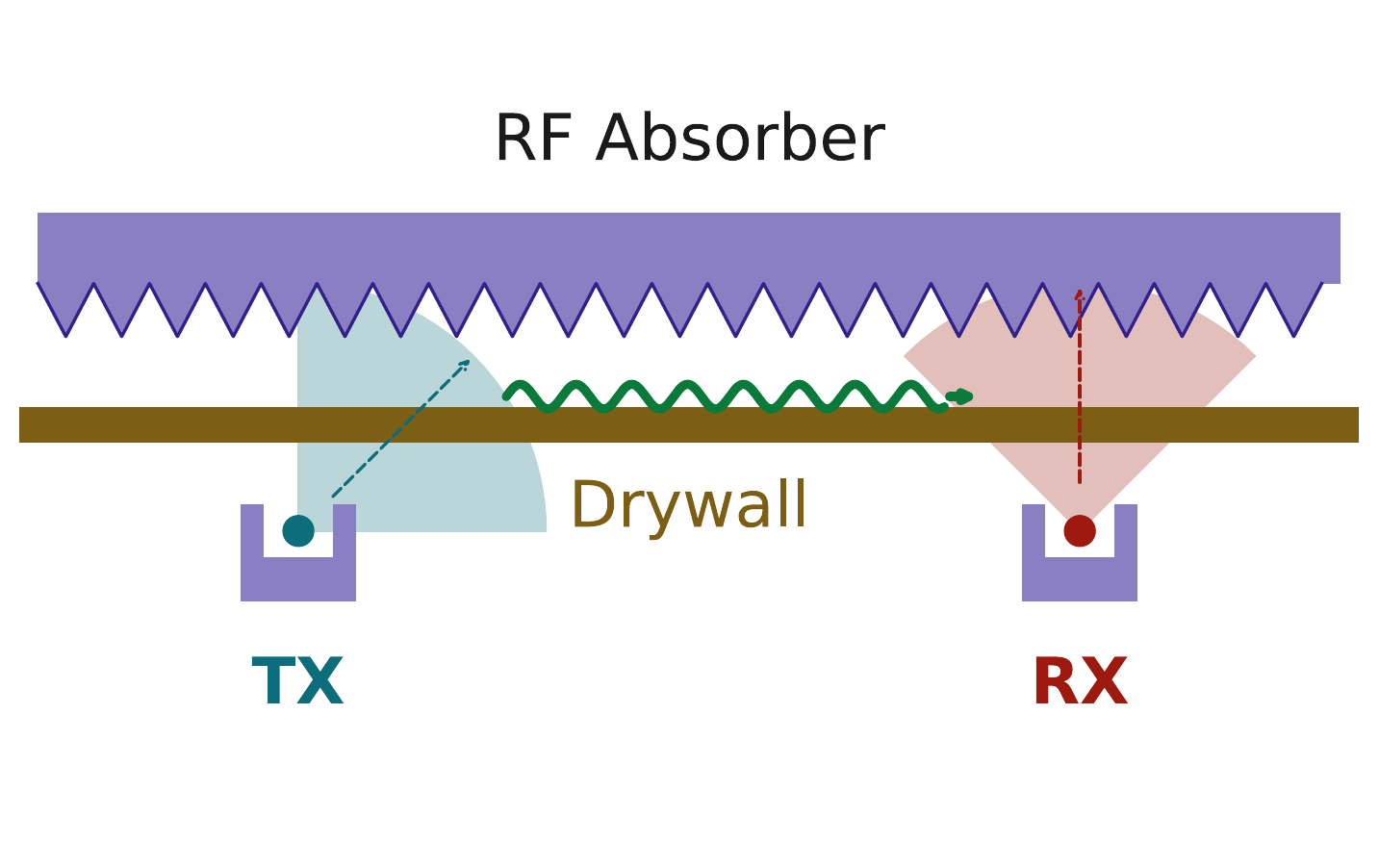}
    \caption{Wall and RF absorber (WA).}
    \label{fig:cfg_wa}
  \end{subfigure}
  \hfill
  \begin{subfigure}[t]{0.29\linewidth}
    \centering
    \includegraphics[width=\linewidth]{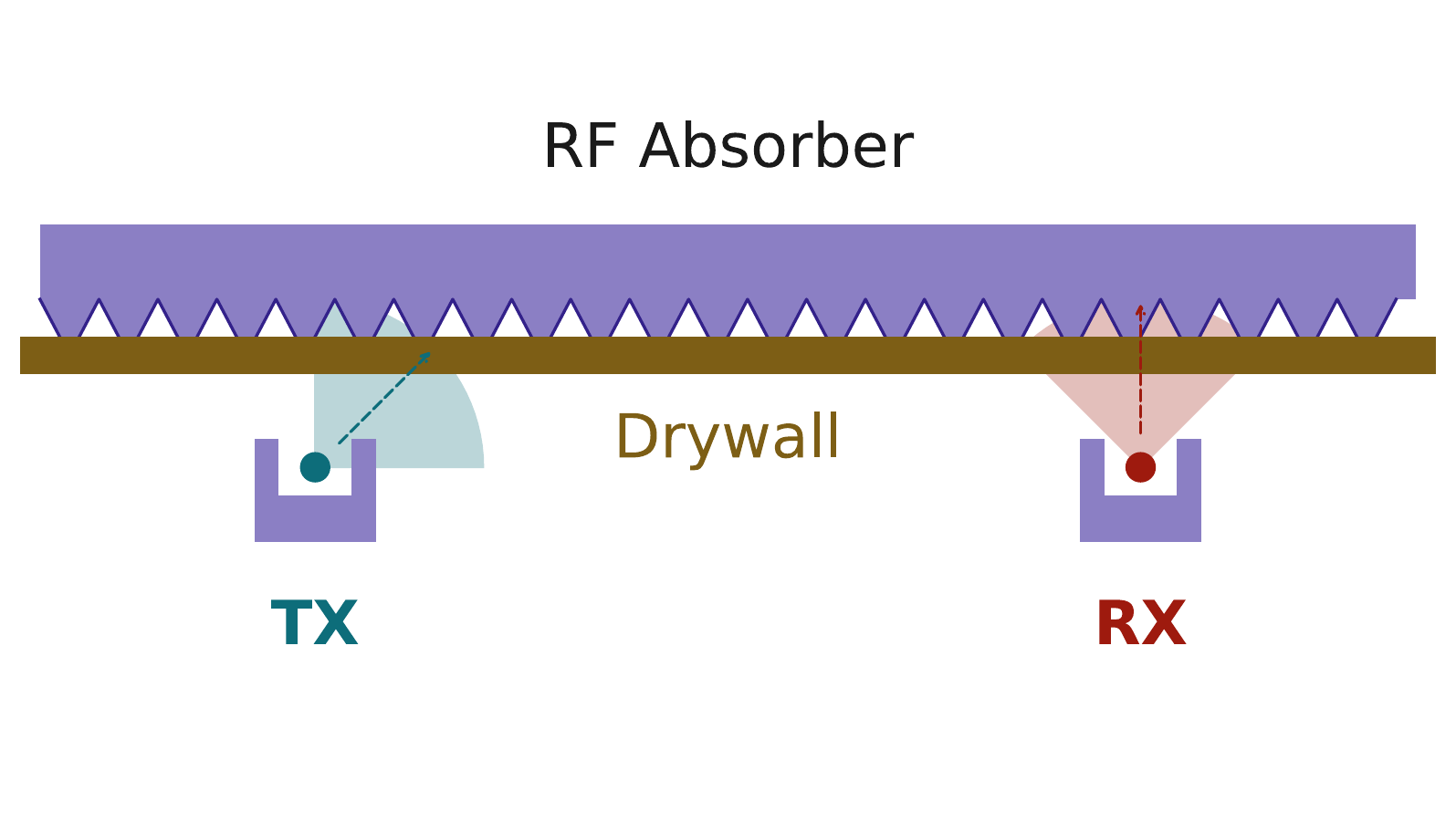}
    \caption{Wall and contacting RF absorber (WCA).}
    \label{fig:cfg_wca}
  \end{subfigure}
  \caption{Measurement setups for three configurations. The lateral wave (green) is absent without drywall~(\subref{fig:cfg_ao}), grazes the far-side air-drywall interface only in the WA case~(\subref{fig:cfg_wa}); and is blocked when the RF absorber contacts the far face~(\subref{fig:cfg_wca}).}
  \label{fig:configs}
\end{figure*}

\begin{figure*}[t]
  \centering
  \begin{subfigure}[t]{0.29\linewidth}
    \centering
    \includegraphics[width=\linewidth]{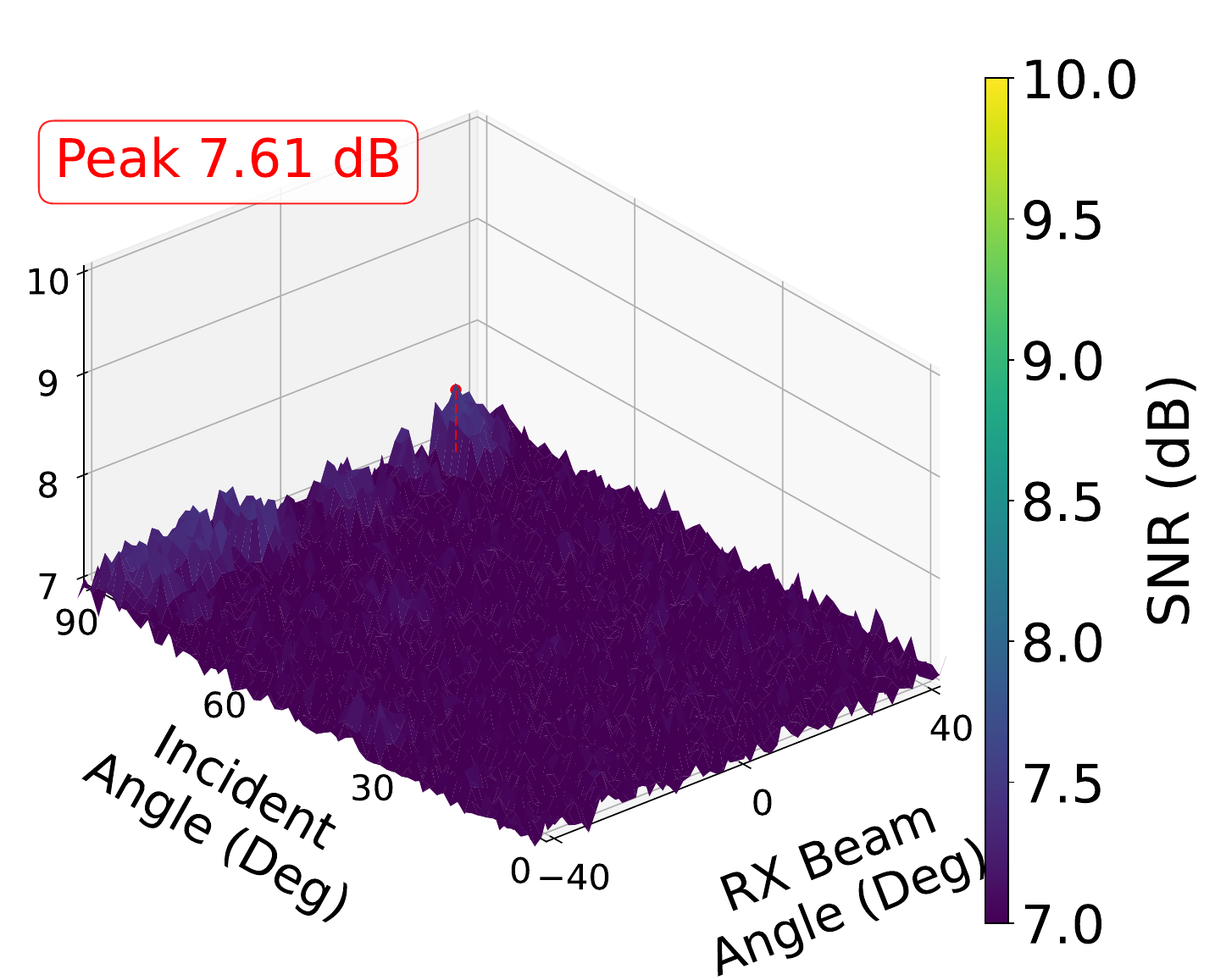}
    \caption{RF absorber only (AO).}
    \label{fig:3d_ao}
  \end{subfigure}
  \vspace{-0.25cm}
  \hfill
  \begin{subfigure}[t]{0.29\linewidth}
    \centering
    \includegraphics[width=\linewidth]{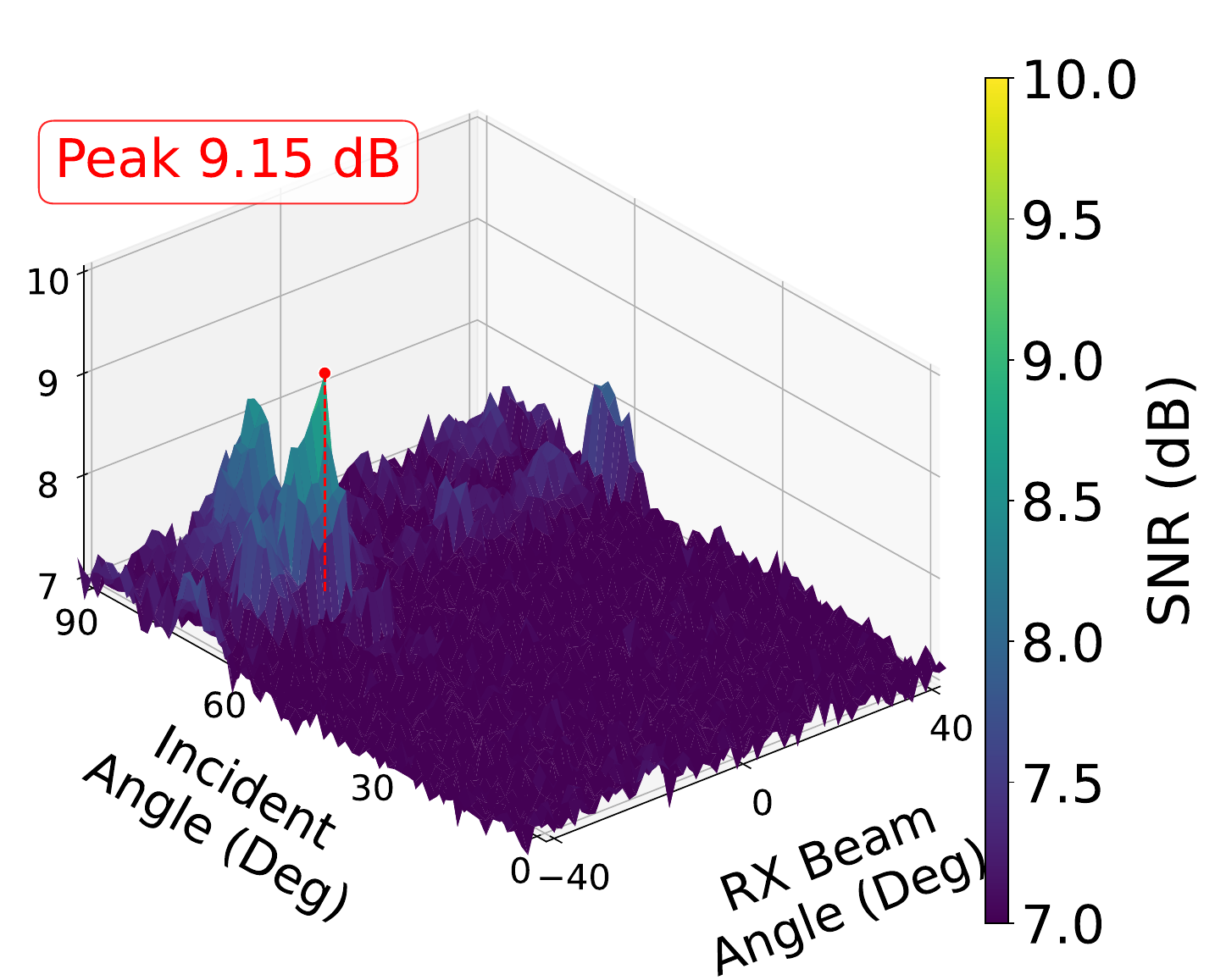}
    \caption{Wall and RF absorber (WA).}
    \label{fig:3d_wa}
  \end{subfigure}
  \hfill
  \begin{subfigure}[t]{0.29\linewidth}
    \centering
    \includegraphics[width=\linewidth]{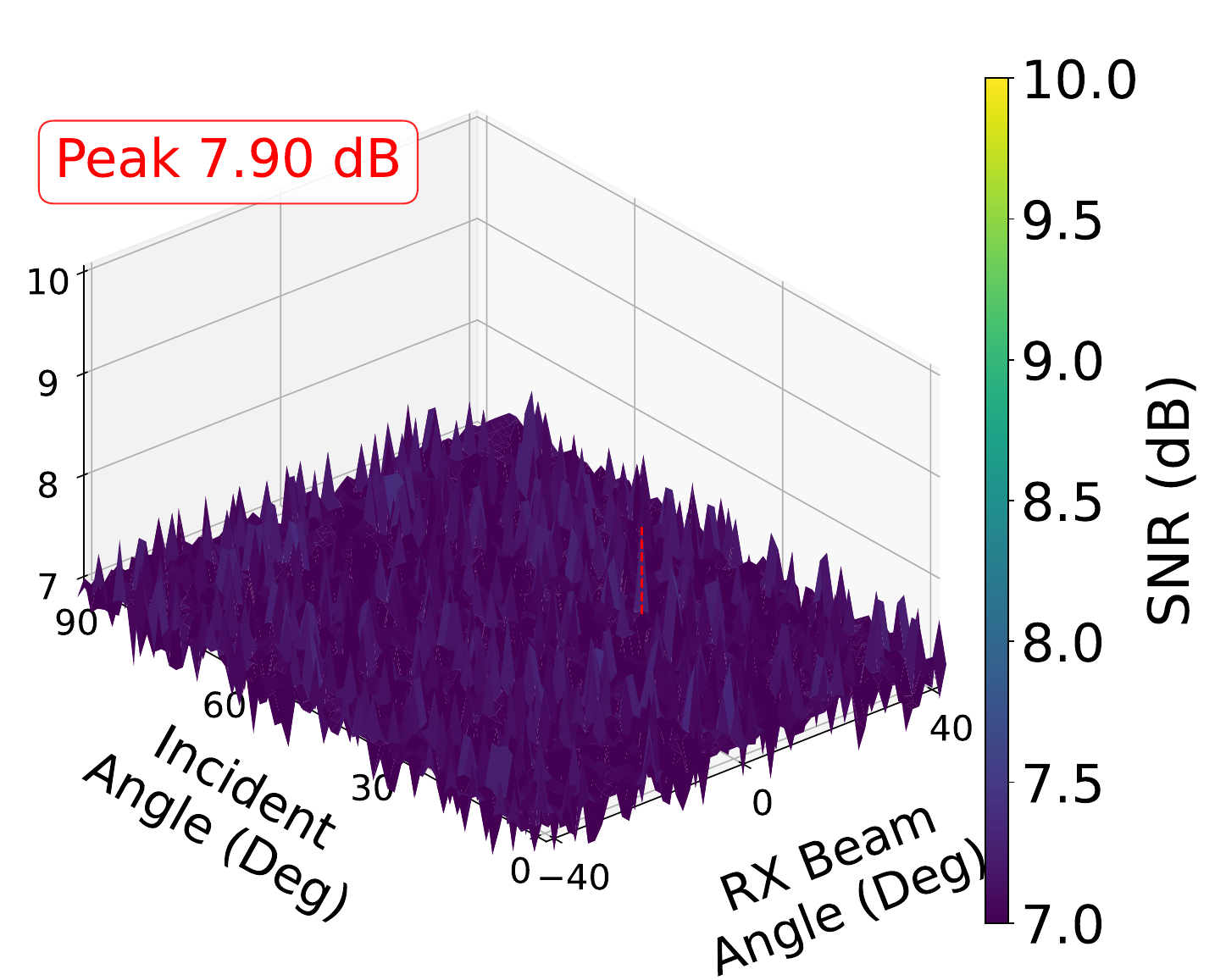}
    \caption{Wall and contacting RF absorber (WCA).}
    \label{fig:3d_wca}
  \end{subfigure}
  \caption{Measured SNR over all TX–RX beam pairs for the three configurations. A high-incidence peak appears only when the drywall is present, and the RF Absorber is placed $5"$ behind it, and vanishes when an RF absorber contacts the far face.}
  \label{fig:3d_heatmaps}
  \vspace{-0.25cm}
\end{figure*}


In the following, we first review the propagation characteristics of building materials. Then, we discuss reconfigurable intelligent surfaces that exploit wall reflections. Next, we cover the background on lateral-wave propagation. Finally, we discuss its use in nanonetworks and the Internet of Paint.

\vspace{-0.1cm}
\subsection{Propagation Characteristics of Building Materials}

Wireless propagation studies treat the wall mainly as an obstacle to be measured.
Measurements at 28, 73, and 91\,GHz report attenuation in common building materials~\cite{hosseini2021attenuation}. Light partitions such as drywall pass the signal, while dense materials block it.
Separate measurements extend penetration loss into the sub-terahertz range for indoor environments~\cite{du2021sub}, and other indoor studies characterize both penetration and reflection at mmWave~\cite{khatun2021penetration}.
At terahertz frequencies, reflections from layered building materials reshape the channel~\cite{jansen2008impact}, and rough surfaces add diffuse scattering~\cite{jansen2011diffuse,piesiewicz2007scattering}.
Standardized models and surveys consolidate these permittivity, reflection, and scattering effects in~\cite{itu2040effects,han2022terahertz}.
This literature quantifies loss, reflection, and scattering, yet it stops short of resolving a wave guided along the wall surface itself.

\vspace{-0.1cm}
\subsection{Intelligent Reflecting Surface/Reconfigurable Intelligent Surfaces}
Since a wall reflects and scatters so strongly, a large body of work turns these reflections into a resource.
Reconfigurable intelligent surfaces (RIS) mount a fabricated array of tunable elements on the wall and reconfigure it to steer the reflected signal around obstacles~\cite{di2020smart,wu2020ris}.
Building such a surface means adding engineered hardware to the wall. It requires a fabricated array of sub-wavelength tunable elements, along with the bias lines, control circuitry, and power needed to reconfigure them.
The approach rests on reflecting energy from the wall, and requires this fabricated, powered surface to be mounted on it.

\vspace{-0.1cm}
\subsection{Background on Lateral-Wave Propagation}
A different mechanism guides energy along an interface rather than reflecting it.
Classical theory describes this lateral wave for a dipole above a conducting half-space~\cite{sommerfeld1909ausbreitung,banos1966dipole} and develops it into closed-form field expressions for communication~\cite{king2012lateral}.
The lateral wave differs from bound surface waves such as the Zenneck wave and the ground wave~\cite{zenneck1907fortpflanzung,norton1937propagation} and from engineered surface plasmons on structured surfaces~\cite{pendry2004mimicking}.
It is the electromagnetic analog of the seismic head wave~\cite{cerveny1971theory}. Every measured instance, though, shares one setting: a thick, low-loss natural medium such as soil~\cite{dong2011channel,salam2019theoretical}, forest~\cite{tamir1967radio}, glacial ice~\cite{evans1963radio,clough1976electromagnetic}, or water~\cite{smolyaninov2018surface} at HF, VHF, or low-GHz frequencies.

\vspace{-0.1cm}
\subsection{Nanonetworks and Internet of Paint}
Recent nanonetwork research has moved the same mechanism toward embedded radios at much higher frequencies.
Graphene plasmonic antennas and terahertz channel models establish communication between nanoscale devices~\cite{akyildiz2008nanonetworks,jornet2013graphene,llatser2013graphene,elayan2017terahertz}, and multi-ray and causal models refine the terahertz channel models~\cite{han2015multiray,tsujimura2018causal}.
The Internet of Paint embeds such radios in wall paint and models the channel as a set of reflected and lateral paths~\cite{wedage2024internet,wedage2025internet}.
Accordingly, the lateral wave along the air-paint interface is identified as the dominant low-loss path as device separation grows, and accordingly, new patch antenna is designed for this setting~\cite{wedage2025antennas}.
This premise rests on theory and simulation where the underlying lateral-wave formulation has been validated only in natural media at lower frequencies, but lateral waves have not been empirically demonstrated for paint or drywall at mmWave frequencies.
Across these threads, it is important to note that direct experimental evidence of a lateral wave at mmWave along an engineered building-material interface has not been provided. Past building-material studies measure loss and reflection~\cite{hosseini2021attenuation,han2022terahertz} but not propagation along the surface. Lateral-wave experiments remain confined to natural media at lower frequencies~\cite{tamir1967radio,dong2011channel,smolyaninov2018surface}, and the 
Internet of Paint (IoP) remains theoretical~\cite{wedage2024internet}. To the best of our knowledge, the first experimental evidence of lateral waves at mmWave frequencies is provided in this paper.

\vspace{-0.15cm}
\section{Overview}
\label{sec:overview}
In this section, we experimentally demonstrate that a lateral wave exists at the air-drywall interface at mmWave frequencies. The transmitter (TX) and receiver (RX) 
are positioned in air on the same side of the sheet, and the lateral wave couples through the drywall to graze the far air-drywall interface before penetrating back into the wall. 
Using a 60\,GHz beamforming module and an standard gypsum drywall sheet, we leverage previously reported drywall dielectric properties to determine the critical angle at which the lateral wave couples to the air-wall interface. We then probe the interface with a bistatic beam sweep over all TX and RX beam pairs and record the signal-to-noise ratio (SNR). We show that a distinct return at high incidence appears only when the drywall is present, matching the grazing geometry of the lateral wave. We obtain evidence that validates the model in Section~\ref{sec:channel_model}.

\subsection{Dielectric Properties of Drywall}
\label{subsec:dielectric}
Gypsum drywall is a low-loss dielectric, which is what makes a lateral wave along its surface plausible in the first place. A 60\,GHz characterization of indoor drywall in~\cite{khatun2021penetration} reports a refractive index of $n_w = 1.88$ and $\mathrm{Im}(n_w) \approx 0.017$. This corresponds to a real relative permittivity $\varepsilon_r = n_w^2 \approx 3.5$. 
The ITU-R~P.2040\cite{itu2040effects} building-material model reports a lower value for generic plasterboard, $\varepsilon_r \approx 2.6$--$2.8$ in the 60\,GHz band ($n_w \approx 1.6$)~\cite{itu2040effects}. Consistent with this low-loss property, mmWave measurement campaigns report only a few decibels of one-way penetration loss through standard $1/2$-in (12.7\,mm) drywall, far below the tens of decibels incurred by denser partitions~\cite{hosseini2021attenuation,du2021sub,khatun2021penetration}. These refractive indices place the drywall-air critical angle, $\theta_c = \arcsin(n_a/n_w)$, at  $32.1^\circ$ for the measured drywall and $38^\circ$ for the ITU plasterboard model.

\vspace{-0.18cm}
\subsection{Lateral Wave Propagation }
A lateral wave is a head wave that travels along the boundary between two media of different refractive indices~\cite{cerveny1971theory} as shown in Fig.~\ref{fig:cfg_wa}. We experimentally establish that such a wave exists at mmWave frequencies along the air-drywall interface, using only a 60\,GHz beamforming module and a sheet of gypsum drywall, and model it in Section~\ref{sec:channel_model}.
We probe the interface between air and drywall with a bistatic beam sweep, repeated in each of the three configurations of Fig.~\ref{fig:configs}.
TX and RX each steer across a codebook of beams spanning $-45^\circ$ to $+45^\circ$ about the antenna boresight.
The incidence angle $\theta_{\mathrm{i}}$ at the wall is the sum of the mechanical TX boresight angle $\theta_{b}$ and the steered beam angle $\theta_{t}$, such that $\theta_{\mathrm{i}} = \theta_{b} + \theta_{t}$.
With the boresight at normal ($\theta_{b} = 0^\circ$), the codebook reaches at most $45^\circ$ of incidence, short of the near-grazing angles at which the lateral wave couples.
Accordingly, we rotate the TX boresight across a set of angles so that the $0^\circ$ beam is incident on the drywall at progressively sharper angles, approaching grazing incidence.
Each TX-RX beam pair yields a mean SNR over 2,000 samples.
From the full sweep, we obtain a two-dimensional SNR heatmap over all beam pairs, resolving received power against incidence angle.
A lateral wave appears in this map as a distinct return at high incidence, near grazing.

We compare the three configurations of Fig.~\ref{fig:configs}, where we isolate the interface response and confirm its origin.
In the experiments, TX and RX are placed in RF-absorber-lined enclosures, leaving only the antenna front face open. The specifically-designed enclosures ensure that back-lobe and side-lobe radiations are suppressed. 
\begin{itemize}
  \item \textbf{RF Absorber Only (AO):} An RF absorber is present without the drywall, which gives the response of a lossy target but no interface (Fig. ~\ref{fig:cfg_ao}).
  \item \textbf{Wall and RF Absorber (WA):} The drywall is present with an RF absorber behind it, which introduces the interface, while the absorber suppresses the back-face reflection (Fig.~\ref{fig:cfg_wa}).
  \item \textbf{Wall and Contacting RF Absorber (WCA):} The drywall is present with the RF absorber pressed against its far face, replacing the low-loss air that guides the lateral wave with an absorbing termination in direct contact with the interface (Fig.~\ref{fig:cfg_wca}).
\end{itemize}

In Figs. \ref{fig:3d_heatmaps}, we show the 3D heatmap of SNR (z-axis) corresponding to incident angle (y-axis) and RX beam angle (x-axis) for each of the configurations in Figs. \ref{fig:configs}. We observe that without the drywall, in the RF absorber-only configuration (Fig.~\ref{fig:cfg_ao}), the SNR stays at $\approx 7.5$\,dB across the measured incident angle range, with no distinct peaks (Fig.~\ref{fig:3d_ao}). Introducing the drywall (Fig.~\ref{fig:cfg_wa}) produces a peak SNR of $9.15$\,dB at high incidence angles (Fig.~\ref{fig:3d_wa}), a feature that is absent when only the RF absorber is present. The peaks appear at the grazing angles, at which the lateral waves propagate, and the absorber behind the drywall suppresses the back-face reflection.
Pressing the RF absorber flush against the far face (Fig.~\ref{fig:cfg_wca}) removes this return from lateral waves entirely (Fig.~\ref{fig:3d_wca}). 
From this observation, we confirm that the lateral wave exists at mmWave frequencies, arising from a plain sheet of drywall and an mmWave beamforming module.
In Section~\ref{sec:channel_model}, we model this lateral wave.

\section{Lateral Wave Propagation Channel Model}
\label{sec:channel_model}
In this section, we model the lateral wave propagation. The lateral wave carries energy between a TX and RX in air on the same side of a wall, where the wave couples through the sheet and grazes along the far-side air-wall interface.
The model builds on the IoP channel formulation~\cite{wedage2024internet}, restructured for terminals that are positioned
in air.
\begin{figure}[t!]
    \centering
    \includegraphics[width=\linewidth]{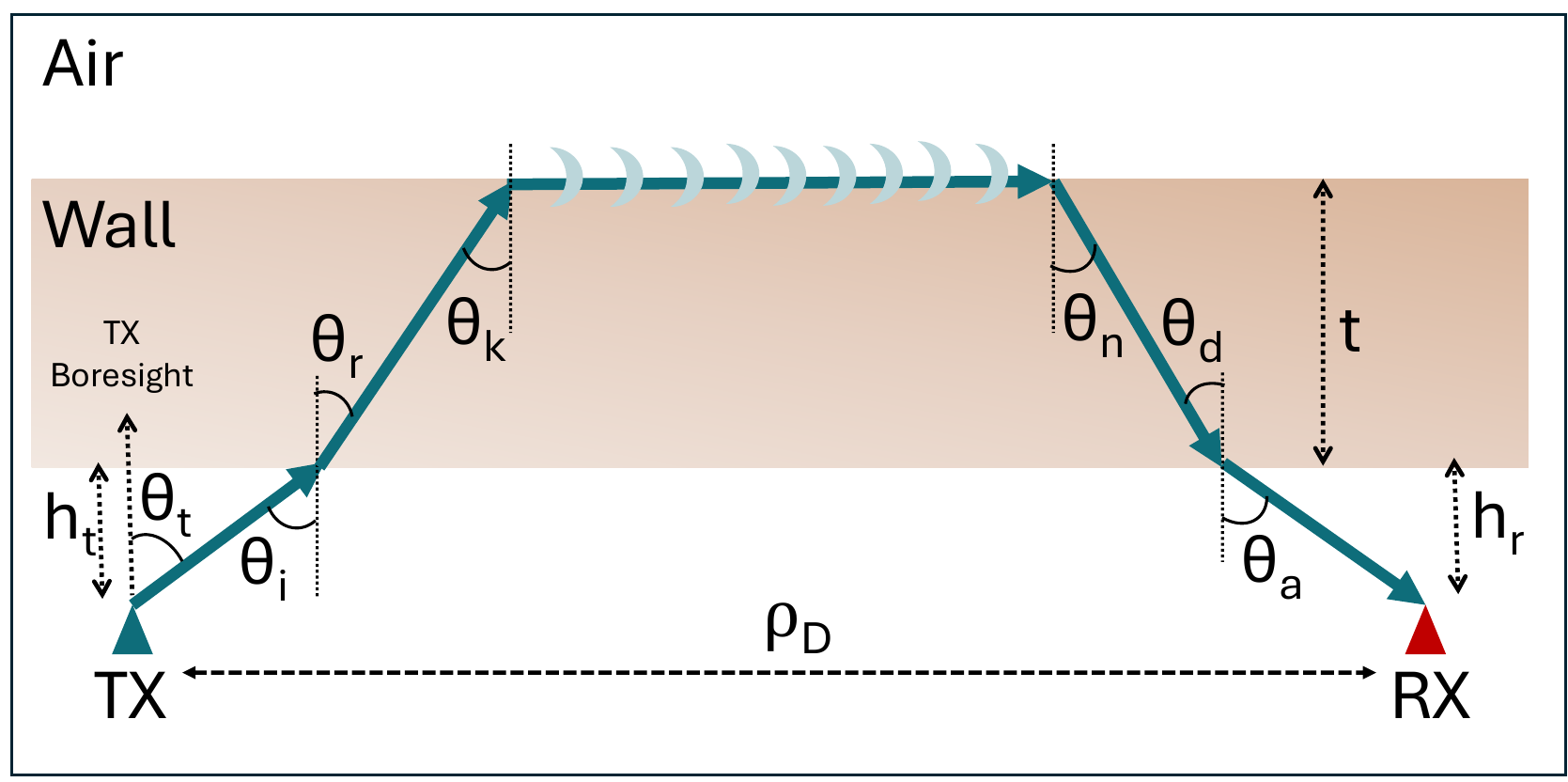}
    \caption{Lateral wave propagation along the wall. The TX ray refracts through the near face and couples at the far-side air-drywall interface at $\theta_k$, travels along the interface over $d_{LW}$, then re-enters the wall at $\theta_{n}$ and exits at $\theta_{d}$.}
    \label{fig:lateral-refracted-wave}
    \vspace{-0.2in}
\end{figure}
\paragraph{\textbf{Geometry and Notation}}
As shown in Fig.~\ref{fig:lateral-refracted-wave}, TX and RX are placed in air on the same side of a wall of thickness $t$, and are separated by a horizontal distance $\rho_D$. Air has refractive index $n_a = 1$, and the wall has $n_w$. The carrier frequency is $f$, the free-space speed is $c$, and the phase speed in the wall is $c_w = c/n_w$. TX and RX are positioned near the face of the wall by the perpendicular distances $h_t$ and $h_r$, respectively. 
On the TX side, the ray leaves TX at the incidence angle $\theta_i$ from the normal to the interface. This incidence angle is $\theta_i = \theta_b + \theta_t$ where $\theta_{b}$ is the mechanical TX boresight angle and $\theta_{t}$ is the steered beam angle. This refracts into the wall at $\theta_r$, reaching the far face at the angle $\theta_k$. On the RX side, the wave refracts into the wall at $\theta_n$, meets the near face at $\theta_{d}$, and refracts back into air at $\theta_{a}$ toward the RX. 
The lateral distance traveled along the interface in air is $d_{LW}$. The lateral wave is formed when $\theta_i$ and $\theta_{a}$ approach $90^\circ$. According to Snell's law, near-grazing incidence refracts the ray to reach the far face at the critical angle, creating a lateral wave at the air-wall surface.

\paragraph{\textbf{Path Segments}}
The lateral wave path decomposes into five segments (Fig.~\ref{fig:lateral-refracted-wave}): (i) an air leg from TX to the near face, (ii) a slab crossing to the far face, (iii) the lateral wave of length $d_{LW}$ along the far interface, and (iv) the mirrored slab crossing on the RX side, and (v) an air leg from the near face to RX. 
Projecting the two slab crossings onto the interface gives the lateral wave leg,
\begin{equation}
  d_{LW} = \rho_D - (h_t + h_r)\tan\theta_i - 2t\tan\theta_c .
  \label{eq:h5}
\end{equation}
where $h_t$ and $h_r$ are TX and RX distance to the wall, $t$, is the wall thickness. For lateral waves to be formed $\theta_k = \theta_n = \theta_c$ with critical angle, $\theta_c$. Furthermore, since TX and RX are both in air, $\theta_i=\theta_a$. Across a distance sweep, $d_{LW}$ differs from $\rho_D$ by a fixed constant.


\paragraph{\textbf{Air Leg Coupling}}
When the incident angle, $\theta_i$, approaches \ang{90}, the air leg contributes to how the lateral wave is formed and received. However, when TX and RX are placed at small distances from the wall, the impacts of the air legs do not change with the TX-RX distance, $\rho_D$. We consider the air leg as excitation factors~\cite{king2012lateral}. Accordingly, we represent this loss as a constant $L_A$. 

\paragraph{\textbf{Wall Crossings}}

The wave crosses the drywall twice, with refraction angles $\theta_r$ and $\theta_d$ at the respective interfaces. The corresponding propagation distances through the drywall are $t/\cos\theta_r$ and $t/\cos\theta_d$, giving a combined path length of $t/\cos\theta_r+t/\cos\theta_d$. Following the two-medium formulation~\cite[Eq.~(10)]{wedage2024internet}, propagation through the wall adds spreading loss determined by the waves' phase velocity in the wall and bulk absorption, where absorption over a path length $x$ contributes $10\log_{10}(e^{x}) = 4.343\,x$~dB:
\begin{equation}
\begin{aligned}
  PL_\mathrm{s} ={}&
    20\log_{10}\!\left[\frac{4\pi f}{c_w}\,
      t\!\left(\frac{1}{\cos\theta_r}+\frac{1}{\cos\theta_d}\right)\right] \\
    &+ 4.343\,K_w(f)\, t\!\left(\frac{1}{\cos\theta_r}
      +\frac{1}{\cos\theta_d}\right).
\end{aligned}
  \label{eq:pl_slab}
\end{equation}

The wall absorption coefficient is determined by the imaginary part of the refractive index, $K_w(f) = (4\pi f/c)\,\mathrm{Im}(n_w)$.
Unlike the embedded-nanodevice IoP formulation in~\cite{wedage2025antennas}, 
where each terminal couples through a depth of dense medium, only the wall thickness $t$ appears here. 

\paragraph{\textbf{Interface transmission}}
The TX and RX side crossings of the wall faces are ordinary refraction events and incur a Fresnel transmission loss. For the perpendicular polarization, the field transmission coefficient, $\mathsf{t}_{aw}$, and the associated power transmittance, $\tau_{aw}$, at the entry face, where the wave crosses from air into the wall at an angle of incidence, $\theta_i$, and angle of refraction, $\theta_r$, are
\begin{equation}
  \mathsf{t}_{aw} =
    \frac{2 n_a \cos\theta_i}{\,n_a\cos\theta_i + n_w\cos\theta_r\,},
  \qquad
  \tau_{aw} =
    \frac{n_w\cos\theta_r}{n_a\cos\theta_i}\,
    \lvert \mathsf{t}_{aw}\rvert^{2},
  \label{eq:fresnelT}
\end{equation}
respectively, where $n_a \sin\theta_i = n_w \sin\theta_r$. At the exit face, the wave crosses from the wall back into air at an angle of incidence, $\theta_{d}$, and angle of refraction, $\theta_{a}$, with $n_w \sin\theta_{d} = n_a \sin\theta_{a}$, giving
$\tau_{wa}$ analogously. The total two-crossing transmission loss is
\begin{equation}
  L_T = -10\log_{10}\!\big(\tau_{aw}\,\tau_{wa}\big).
  \label{eq:LT}
\end{equation}
As the incidence angle approaches grazing, the plane wave transmittance falls to zero, and $L_T$ diverges. This does not mean the interface blocks the wave. Rather, it indicates that the lateral wave is formed as a head wave rather than by the plane wave refraction described by the Fresnel equations. At the critical angle, the transmittance is small but finite, so $L_T$ stays bounded.

\paragraph{\textbf{Interface Leg}}

The leg along the far interface carries two distance-dependent losses. The first is spreading. The spherical-segment model in~\cite{wedage2024internet} assigns free-space spreading to the air segment, at $20$\,dB per decade. On the other hand, the asymptotic field over a dielectric half-space decays as $1/d_{LW}^{2}$ in amplitude, i.e., $40$\,dB per decade~\cite{king2012lateral,salam2019theoretical}. Therefore, we keep the spreading exponent, $n_{LW}$, general. The second loss is exponential attenuation. One part is the molecular absorption of air, $K_\mathrm{a}(f)$, computed from ITU-R~P.676 \cite{itu676attenuation} and is of the order of $10$\,dB/km at 60\,GHz~\cite{wedage2024internet,itu676attenuation}. The other part is leakage. Since the sheet has finite thickness, the grazing wave continuously re-radiates through the slab into the air on both sides. This leakage, absent from the two-half-space theory in which the dense medium is infinitely deep, drains the interface wave at a constant rate along the surface. We model it with a leakage coefficient, $\alpha_L$, so that the interface wave is multiplied by an additional factor, $e^{-\alpha_L d_{LW}}$, equivalent to a loss of $4.343\,\alpha_L$~dB per meter.
With respect to the reference distance, $d_0$, the interface leg loss is
\begin{equation}
\begin{aligned}
  PL_i(d_{LW}) ={}
    & 20\log_{10}\!\left(\frac{4\pi f\, d_0}{c}\right)
      + 10\, n_{LW}\log_{10}\!\left(\frac{d_{LW}}{d_0}\right) \\
    & + 4.343\,\big(K_a(f) + \alpha_L\big)\, d_{LW} .
\end{aligned}
  \label{eq:pl_int}
\end{equation}

\paragraph{\textbf{Total Path Loss and Calibration}}
Adding the segment losses gives the lateral wave path loss:
\begin{equation}
  PL_{LW}(\rho_D) =
    PL_s + PL_i(d_{LW}) + L_T + L_A ,
  \label{eq:pl_lw_gen}
\end{equation}
with $d_{LW}$ from~\eqref{eq:h5}. Setting $n_{LW} = 2$ and $\alpha_L = 0$ recovers the spherical segment model of~\cite{wedage2024internet}, while setting $n_{LW} = 4$ with $\alpha_L = 0$ recovers the half-space asymptote of \cite{king2012lateral}.
\paragraph{\textbf{Distance Scaling}}
We analyze the impact of TX--RX distance on the lateral wave path loss as $\Delta PL_{LW}(\rho_D) = PL_{LW}(\rho_D) - PL_{LW}(\rho_{D\,0})$. At a fixed frequency, only the interface leg varies with distance. The geometric offset in~\eqref{eq:h5} is a fixed constant, so $d_{LW}$ differs from $\rho_D$ by a constant, and the distance-independent terms of~\eqref{eq:pl_lw_gen}, namely the slab crossings, the interface reference, $L_T$, and $L_A$, form a fixed offset. Taken relative to a reference distance, $\rho_{D\,0}$, this offset cancels, and the path loss grows with distance as:
\begin{equation}
\begin{aligned}
  \Delta PL_{LW}(\rho_D) ={}& 4.343\,(K_a(f) + \alpha_L)\,(\rho_D - \rho_{D\,0}) \\
  &+ \begin{cases}
      20\log_{10}(\rho_D/\rho_{D\,0}), & n_{LW} = 2, \\
      40\log_{10}(\rho_D/\rho_{D\,0}), & n_{LW} = 4.
    \end{cases}
\end{aligned}
  \label{eq:pl_dist}
\end{equation}
The exponent sets the slope, $20$\,dB per decade in the free-space case and $40$\,dB per decade in the head-wave case, and the leakage adds a term linear in $\rho_D$.

\paragraph{\textbf{Frequency Scaling}}
We analyze the impact of operation frequency on the lateral wave path loss as $\Delta PL_{LW}(f) = PL_{LW}(f) - PL_{LW}(f_0)$, with respect to a reference frequency, $f_0$. Using~\eqref{eq:pl_lw_gen}, the relative path loss is represented as
\begin{equation}
\begin{aligned}
  \Delta PL_{LW}(f) ={}& 40\log_{10}\!\left(\frac{f}{f_0}\right)
    + 4.343\,\frac{2t\,K_w(f_0)}{\cos\theta_k}\!\left(\frac{f}{f_0} - 1\right) \\
  &+ 4.343\,\big(K_a(f) - K_a(f_0)\big)\,d_{LW},
\end{aligned}
  \label{eq:pl_freq}
\end{equation}
which removes all frequency-independent terms. The first term in~\eqref{eq:pl_freq} represents the spreading dependence that scales with $f^4$ and the second term is the frequency-dependent loss through the drywall material, which scales linearly with $f$. The final term provides a small correction due to atmospheric absorption. It can be observed that $L_T$ in~\eqref{eq:LT} is considered frequency-independent because the wall refractive index is assumed to be dispersion-free~\cite {itu2040effects,khatun2021penetration}, and $L_A$ in~\eqref{eq:pl_lw_gen} is also frequency-independent. An important observation from~\eqref{eq:pl_freq} is that the first term dictates an effective frequency exponent of $n_f=4$ due to spreading, compared with $n_f=2$ for free-space propagation, which we test experimentally in Section~\ref{sec:evaluation}.
\vspace{-0.15cm}
\section{Model Comparison and Verification}
\label{sec:evaluation}
In this section, we experimentally validate the lateral wave channel model presented in Section~\ref{sec:channel_model}. We first describe the experimental setup, followed by a comparison of the model with measured path loss under frequency and distance scaling and an analysis of the resulting path-loss exponents.
\subsection{Experimental Setup}
 \begin{figure}
     \centering
     \includegraphics[width=0.9\linewidth]{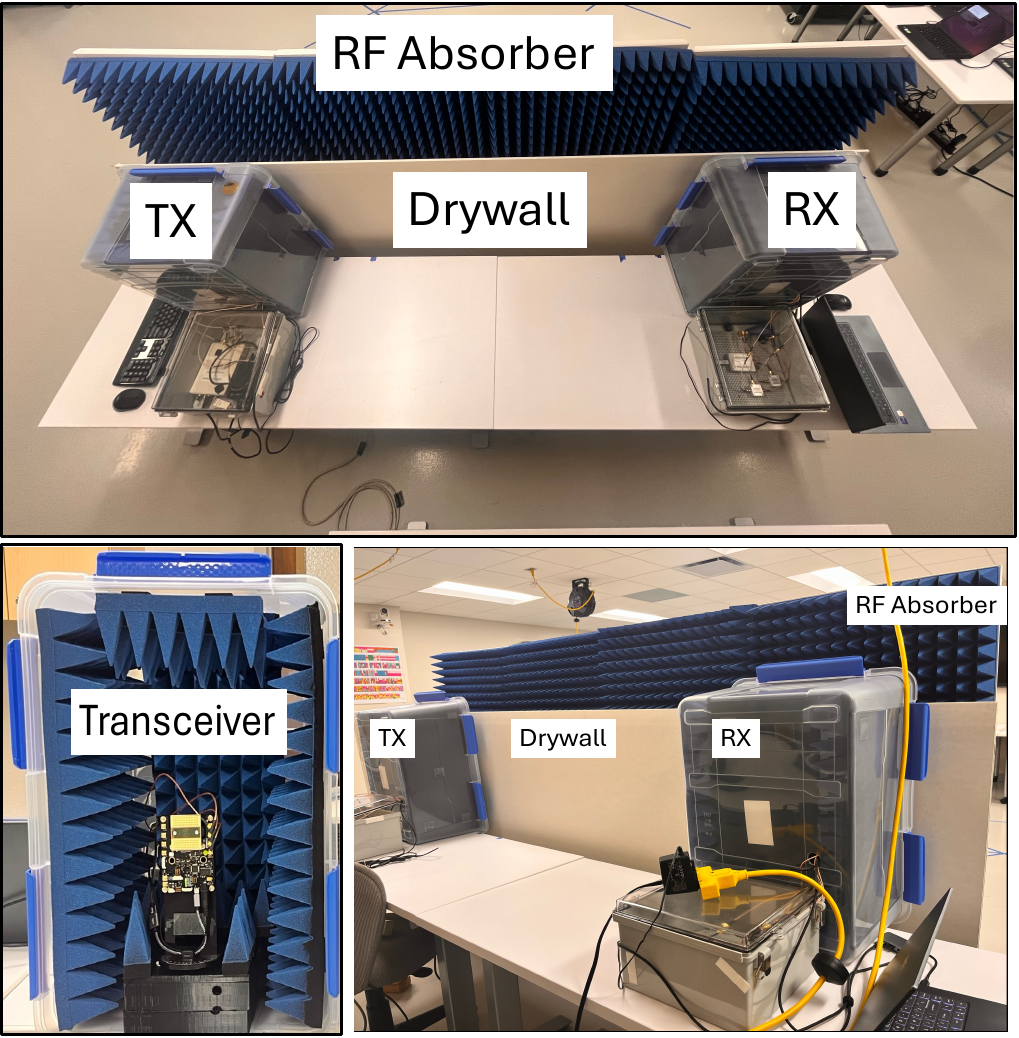}
     \caption{Experimental setup: TX and RX in air on one side of the drywall, with an absorber behind it.}
     \label{fig:setup}
     \vspace{-0.15cm}
 \end{figure}
We show the experimental setup in Fig.~\ref{fig:setup}. We use an ordinary gypsum drywall sheet as a realization of the generic wall modeled in Section~\ref{sec:channel_model}. Two Sivers Semiconductors
EVK 60\,GHz beamforming transceivers \cite{sivers_evk06005} are used as TX and RX, which are positioned in air on the same side of the drywall sheet of size $8$\,ft $\times 4$\,ft and are separated by the distance, $\rho_D$. This transceiver has a $6^\circ$ $3$\,dB-beamwidth. Each transceiver is placed in an enclosure lined with an RF absorber that leaves only the antenna front face open, so that side and back lobe radiation is suppressed and only the main lobes are incident on the drywall. RF absorbers are placed $5$\,in behind the far face of the drywall. This wall and RF absorber (WA) configuration suppresses the reflections behind the far face of the drywall while leaving the far air-drywall interface intact, so that the lateral wave can propagate.

The wave couples through the drywall and lands on the far interface as modeled in Section~\ref{sec:channel_model}. 
TX and RX steer their beam across a codebook spanning $-45^\circ$ to $+45^\circ$ relative to the boresight. We set the TX boresight to $-45^\circ$ relative to the normal of the drywall, such that the beam-steering range corresponds to incident angles from $0^\circ$ to $90^\circ$ with respect to the normal of the drywall. We additionally rotate the TX boresight from \ang{0} to \ang{90}, with intermediate angles of \ang{30}, \ang{45}, \ang{60}, \ang{70}, \ang{75}, and \ang{85}, such that the $0^\circ$ beam is incident on the drywall at progressively sharper angles. These configurations allow the TX beam to reach the near-grazing angles to create the lateral waves which propagates to RX.

We report results averaged over five measurements. To characterize path loss, we focus on $75^\circ-90^\circ$ incidence angles and, for configuration, select the beam pair that yields the strongest received signal. Path loss is then expressed relative to a reference operating point i.e., the lowest frequency ($58$~GHz) for the frequency sweep, and the shortest separation ($\rho_{D\,0} = 1.02$~m) for the distance sweep. The experimental path loss exponent, $n_{\mathrm{exp}}$, is obtained by a least-squares fit of the measured path loss against $\log_{10}$ of frequency or distance, following the $10\,n\log_{10}(\cdot)$ model.
\begin{table}[ht!]
  \centering
  \caption{System and Experimental Parameters.}
  \label{tab:eval_params}
  \begin{tabular}{lll}
    \hline
    Symbol & Value & Unit \\
    \hline
    $f$ & $58$-$69$ & GHz \\
    $c$ & $3.0\times10^{8}$ & m/s \\
    $n_w$ & $1.88$ & unitless \\
    $\mathrm{Im}(n_w)$ & $0.017$ & unitless \\
    $c_w = c/n_w$ & $1.594\times10^{8}$ & m/s \\
    $\theta_k = \theta_n$ & $32.1$ & deg \\
    $t$ & $12.7\times10^{-3}$ & m \\
    $2t/\cos\theta_k$ & $30.0\times10^{-3}$ & m \\
    $K_w(f_0)$ & $41.3$ & Np/m \\
    $K_a(f_0)$ & $1.7\times10^{-3}$ & Np/m \\
    $\alpha_L$ (baseline) & $0$ & Np/m \\
    \hline
  \end{tabular}\vspace{-0.2in}
\end{table}
\subsection{Evaluations}
\begin{figure}[t]
  \centering
  \includegraphics[width=0.9\linewidth]{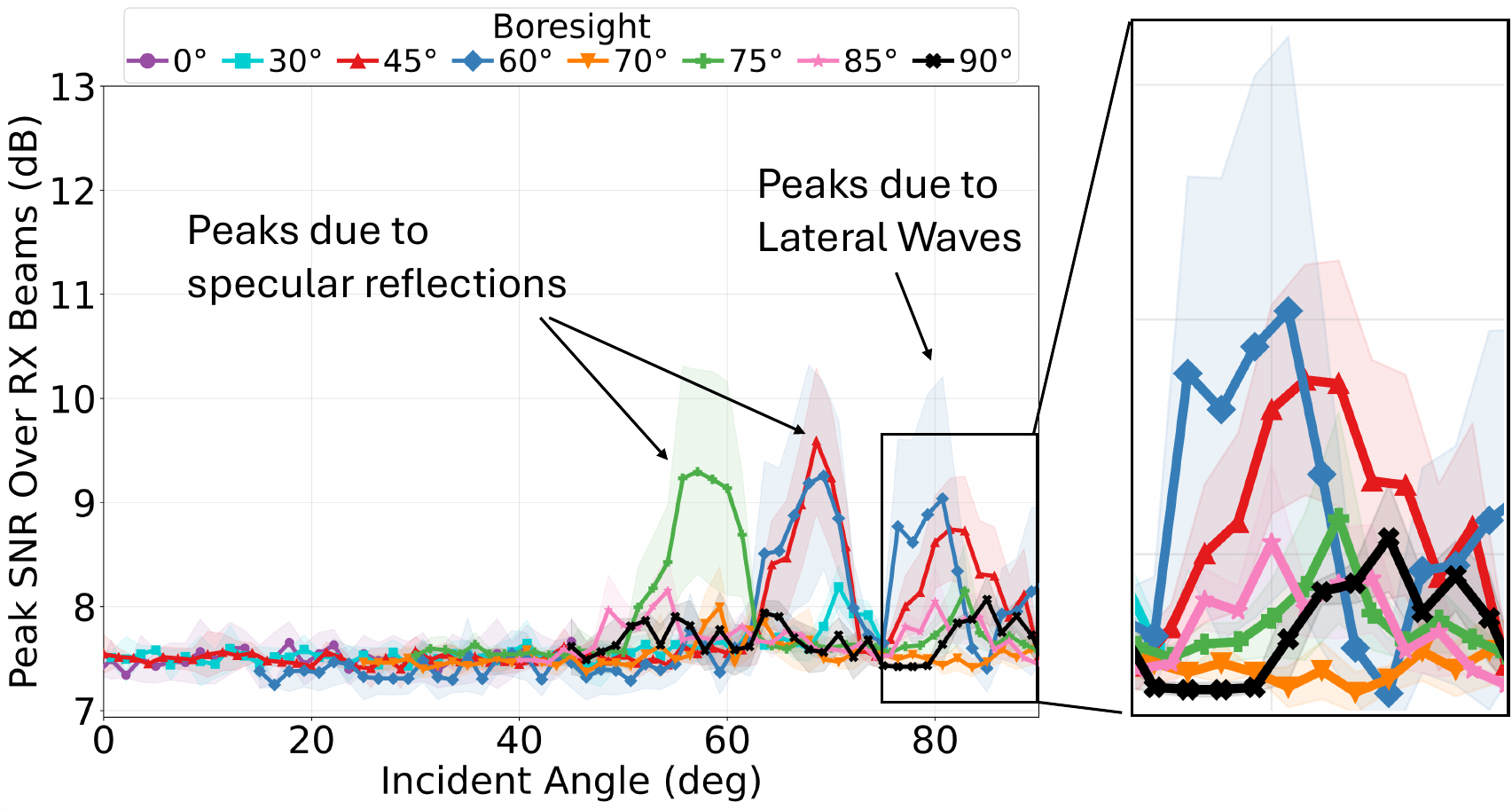}
  \caption{Peak SNR over RX beams versus incidence angle (WA, 60\,GHz).
  SNR peaks near grazing incidence, where the lateral wave forms. 
  Shaded regions represent the standard deviation across five measurements, evaluated at each incidence angle.}
  \label{fig:snr_incident}
  \vspace{-0.30cm}
\end{figure}
\begin{figure*}[t]
  \centering
  \begin{subfigure}[t]{0.28\linewidth}
    \centering
    \includegraphics[width=\linewidth]{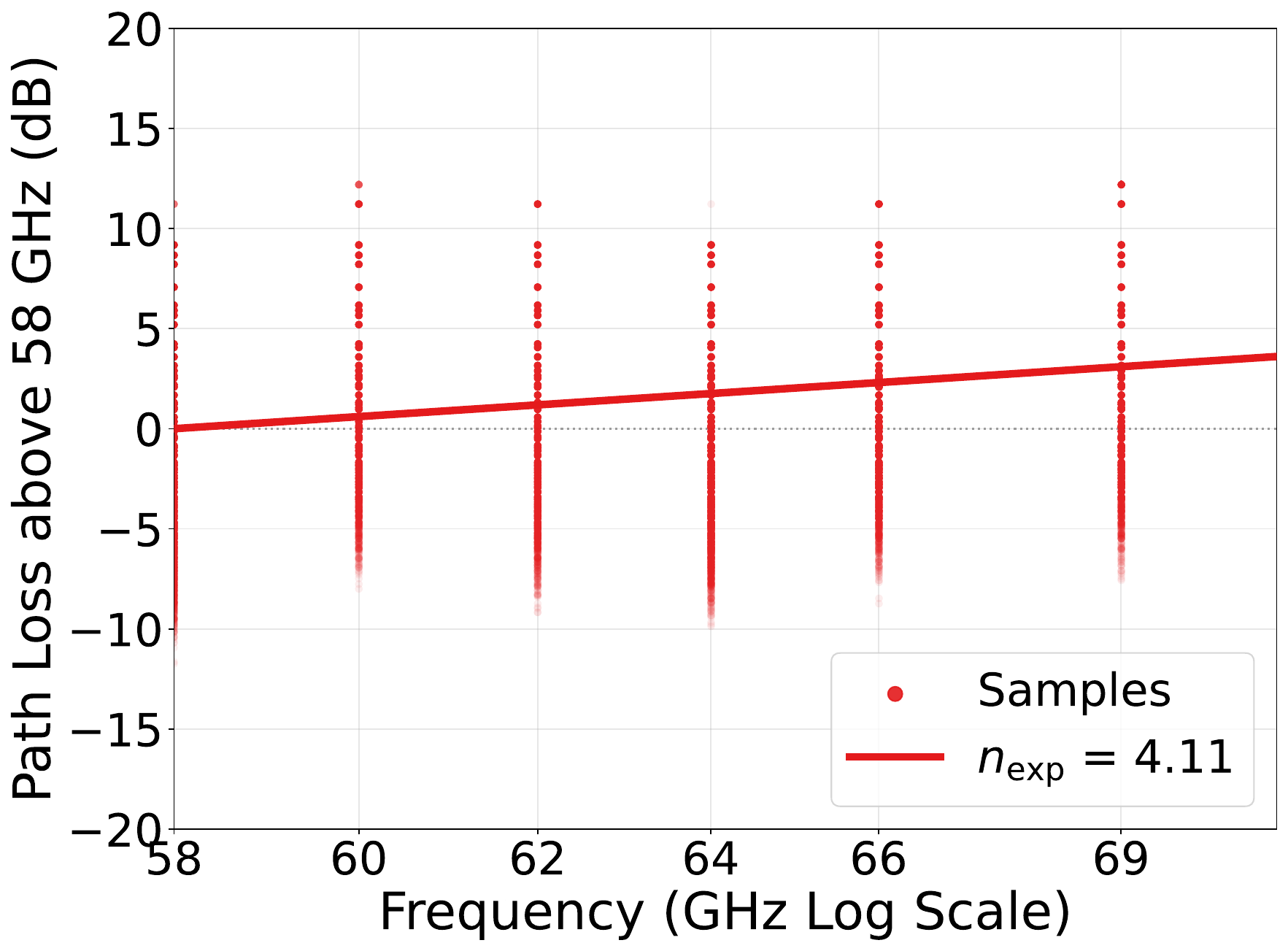}
    \caption{Boresight $45^\circ$.}
    \label{fig:pl_45}
  \end{subfigure}
  \hfill
  \begin{subfigure}[t]{0.28\linewidth}
    \centering
    \includegraphics[width=\linewidth]{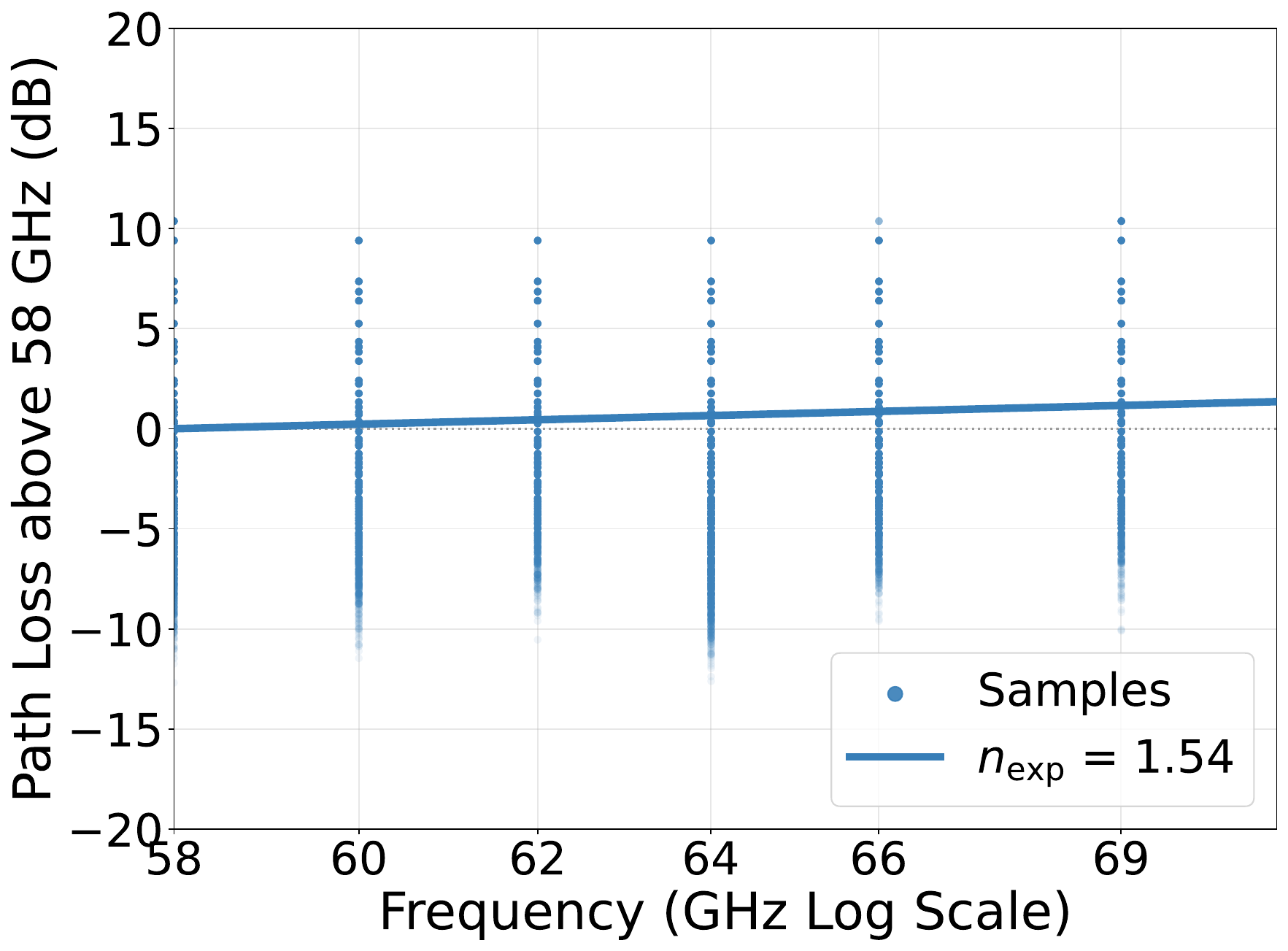}
    \caption{Boresight $60^\circ$.}
    \label{fig:pl_60}
  \end{subfigure}
  \hfill
  \begin{subfigure}[t]{0.28\linewidth}
    \centering
    \includegraphics[width=\linewidth]{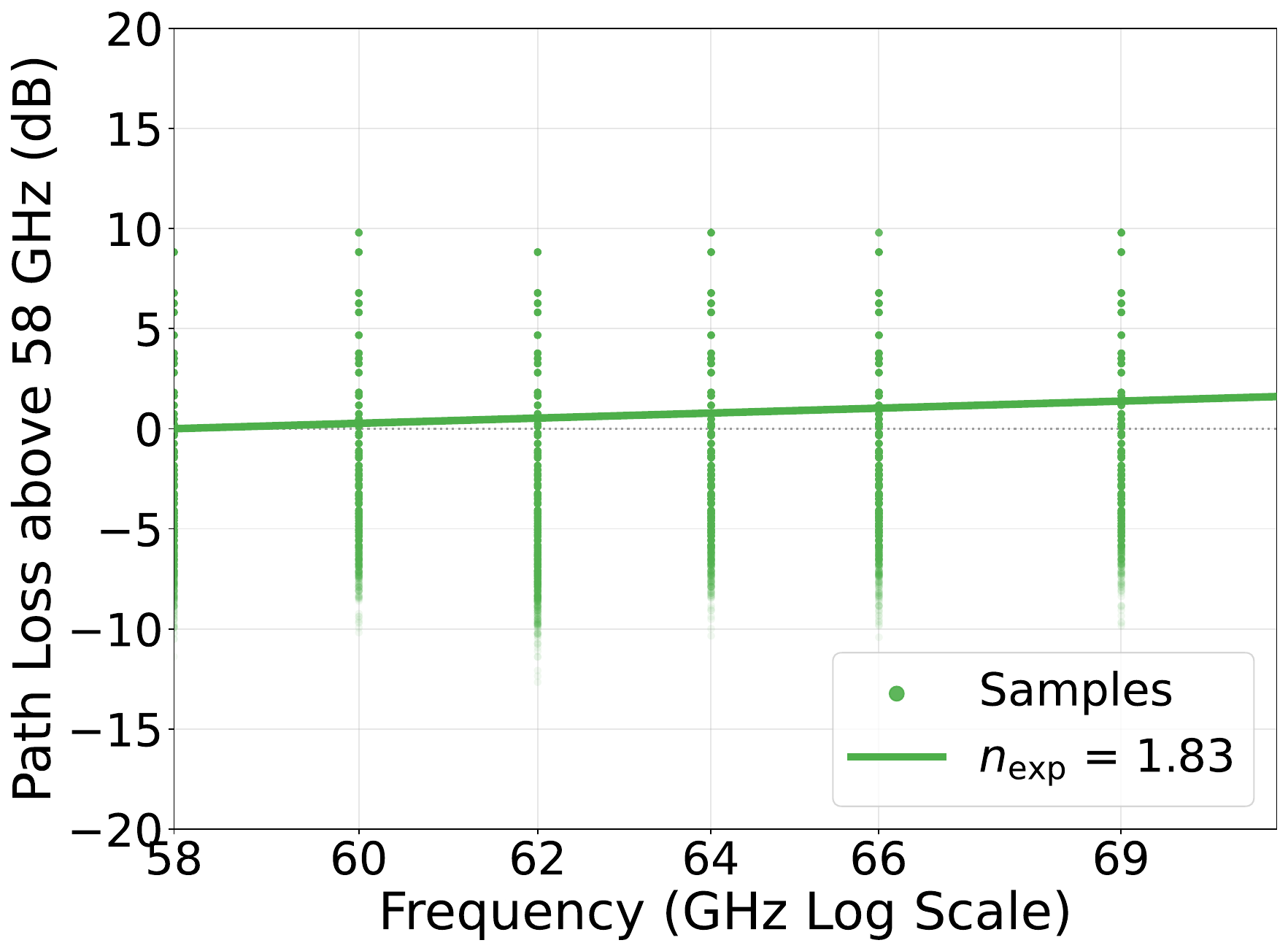}
    \caption{Boresight $75^\circ$.}
    \label{fig:pl_75}
  \end{subfigure}
  \caption{Measured path loss versus frequency for the WA configuration at boresight angles (a)$45^\circ$, (b)$60^\circ$, and (c)~$75^\circ$, over the $58$--$69$\,GHz sweep.}
  \label{fig:pathloss_freq}
  \vspace{-0.15cm}
\end{figure*}
In this section, we evaluate the model in Section~\ref{sec:channel_model} in two stages: We first confirm that the high-incidence return is caused by lateral waves. It appears at near-grazing angles, where the refracted ray reaches the far face at the critical angle. We then compare the relative path loss $\Delta PL_{LW}$ as it varies with frequency and distance against the measurement. 
In Table~\ref{tab:eval_params}, we list the experiment parameters. The drywall index $n_w = 1.88$ and $\mathrm{Im}(n_w) \approx 0.017$ are based on the 60\,GHz characterization in~\cite{khatun2021penetration}.
\paragraph{\textbf{Critical Angle at 60\,GHz}} 
In Fig.~\ref{fig:snr_incident}, we plot the peak SNR over the RX beams against the incident angle at 60\,GHz. Each curve corresponds to one mechanical boresight, averaged over five measurements. The shaded region represents the standard deviation across the five trials, evaluated at each incidence angle. Below \ang{50}, every boresight stays around $7.5$\,dB.
Above \ang{50}, three distinct peaks are observed. The first peak is observed at \ang{57} for the \ang{75} boresight. The second peak is at \ang{69} incident angle for the \ang{45} and \ang{60} boresights. Finally, we observe peaks at \ang{81}-- \ang{83} for five boresights (\ang{45}, \ang{60}, \ang{75}, \ang{85}, \ang{90}).

From the measurement geometry, the first-order specular reflection for the $1.83$\,m ($72$\,in) TX--RX separation reaches RX at an incidence angle of $85.24^\circ$. This implies a grazing angle ~\cite{king2012lateral, clough1976electromagnetic} of $4.76^\circ$ relative to the wall surface. At a high incidence angle of $84.5^\circ$, the $6^\circ$ beamwidth of our transceivers projects the incident beam over $82^\circ$--$88^\circ$ incidence. Part of the beam reaches within $2^\circ$ of grazing angle. The lateral wave propagates in this near-grazing range. Since the beam codebook is discrete, the strongest measured response occurs at beam angles of $81^\circ$-- $83^\circ$, whose \ang{6} beamwidth overlaps the near-grazing angular region required for lateral wave formation. 

\begin{figure*}[t!]
  \centering
  \begin{subfigure}[ht!]{0.3\linewidth}
    \centering
    \includegraphics[width=\linewidth]{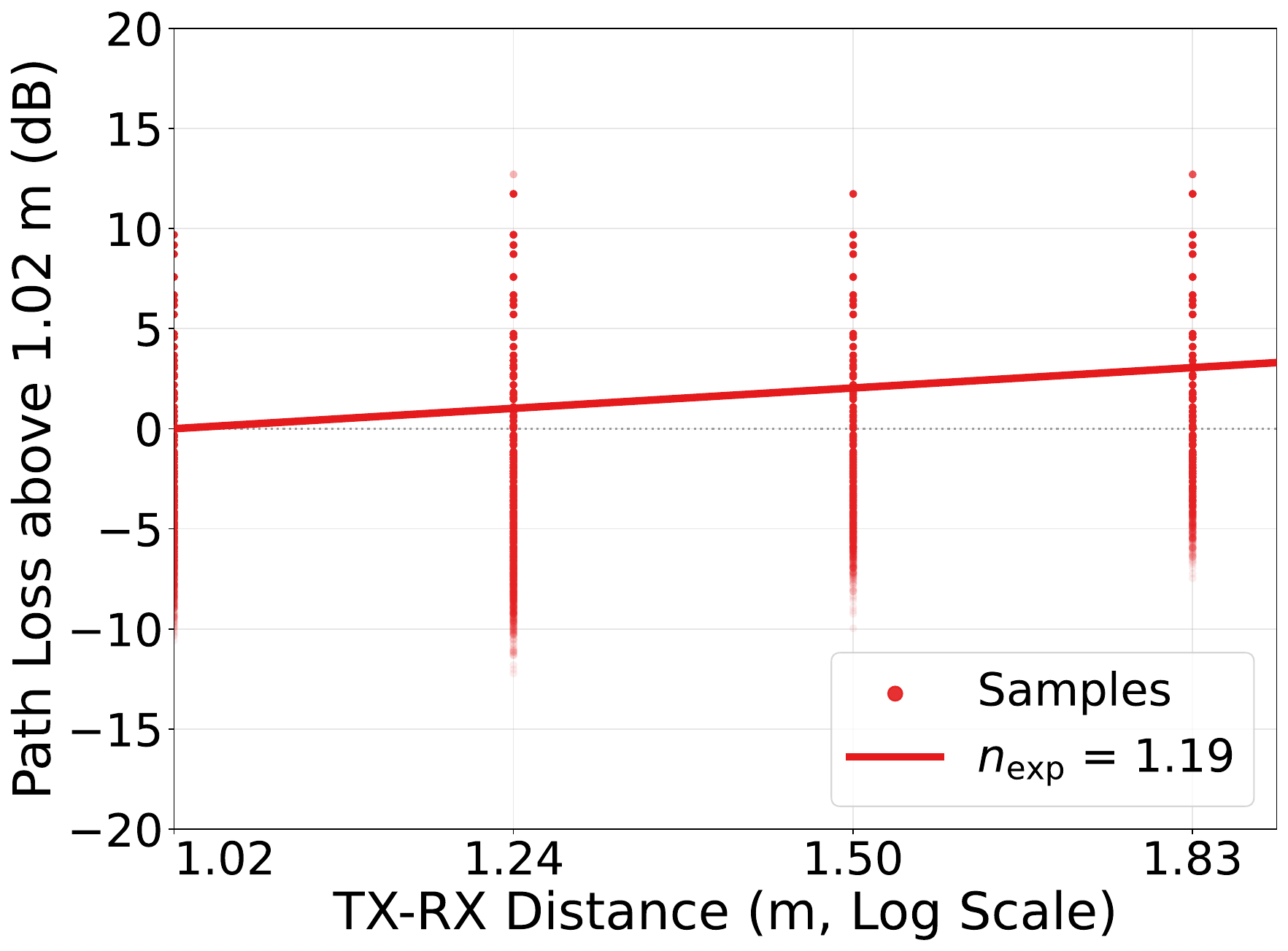}
    \caption{Boresight $45^\circ$.}
    \label{fig:pd_45}
  \end{subfigure}
  \hfill
  \begin{subfigure}[ht!]{0.3\linewidth}
    \centering
    \includegraphics[width=\linewidth]{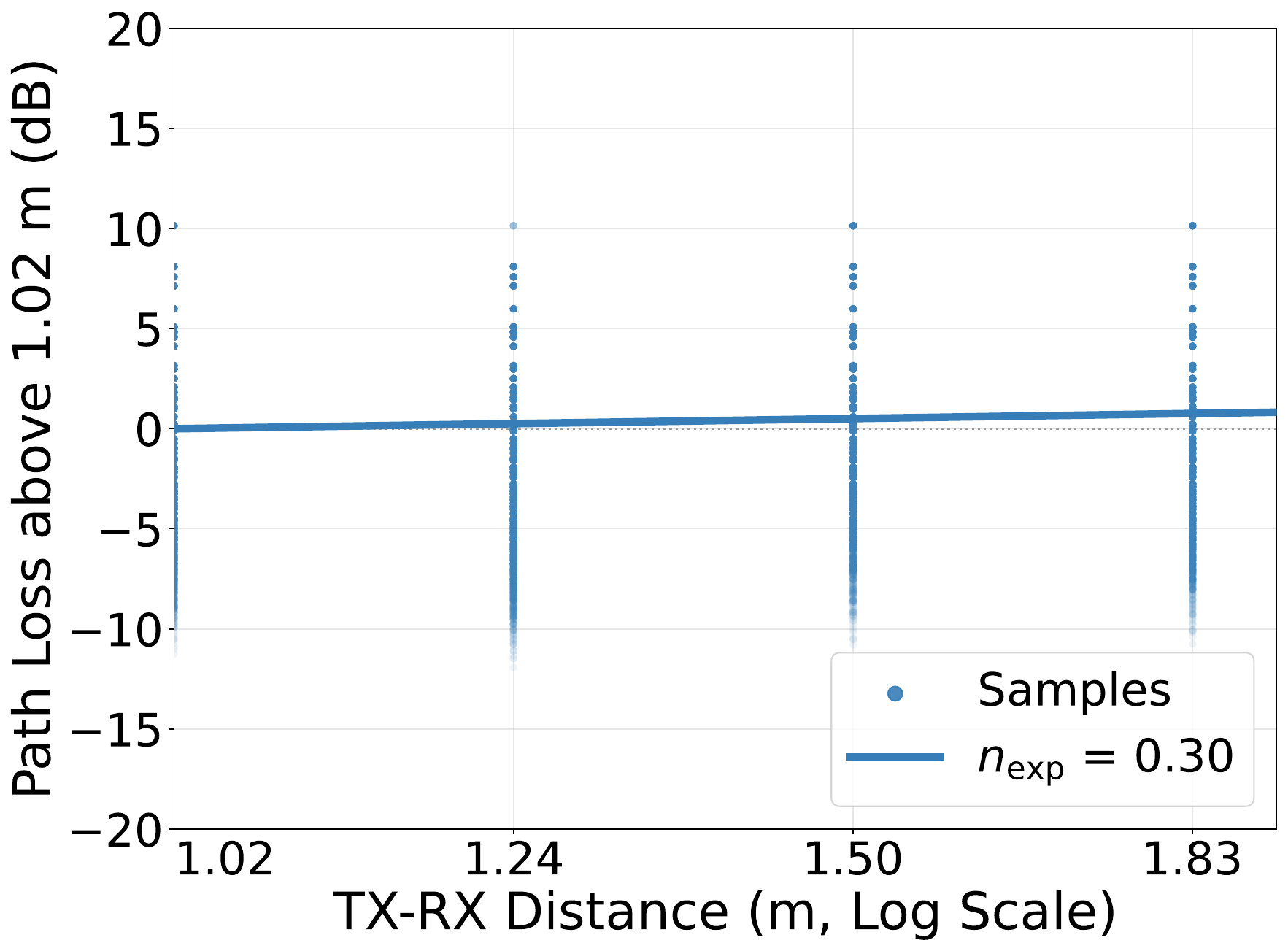}
    \caption{Boresight $60^\circ$.}
    \label{fig:pd_60}
  \end{subfigure}
  \hfill
  \begin{subfigure}[ht!]{0.3\linewidth}
    \centering
    \includegraphics[width=\linewidth]{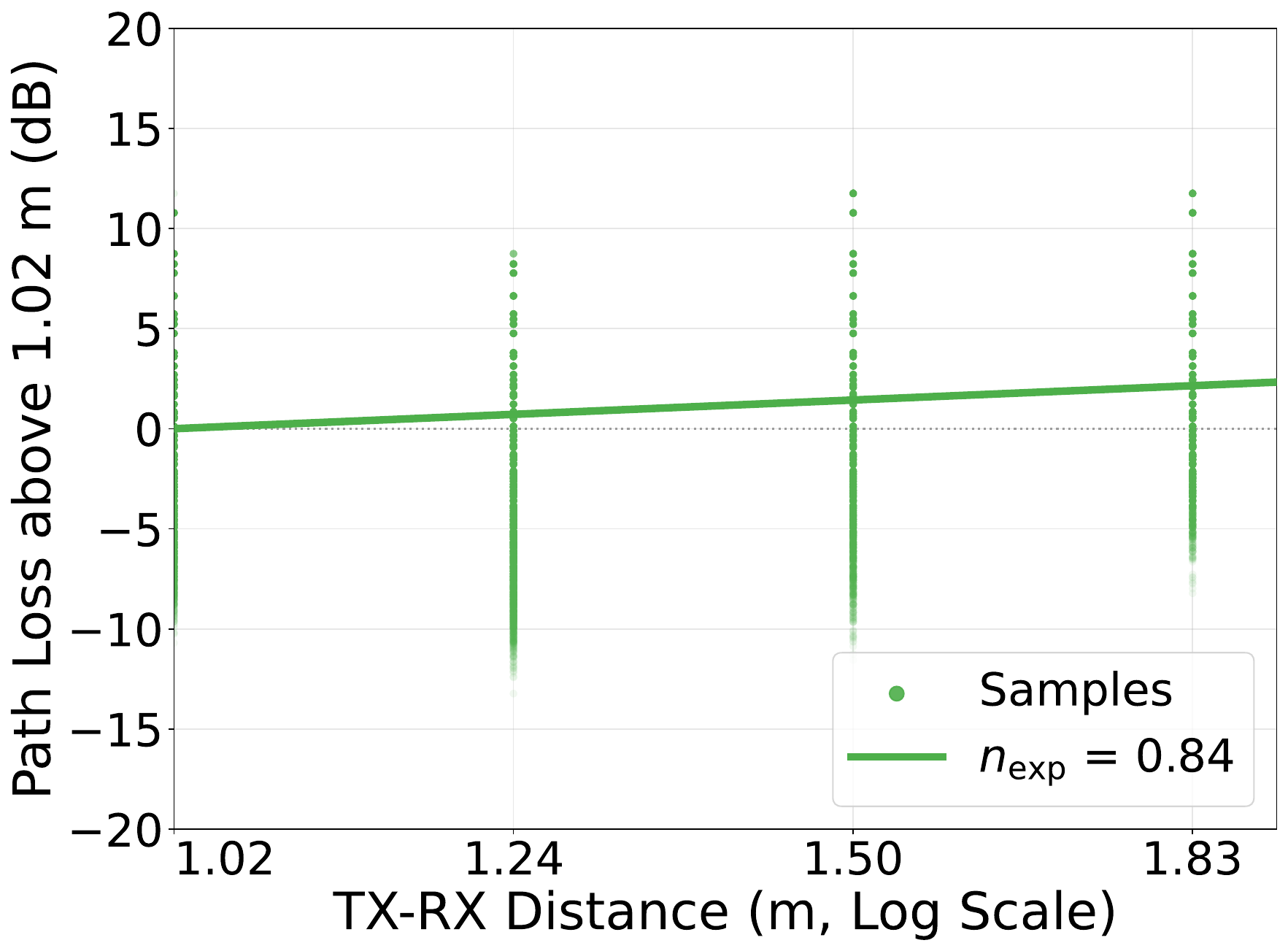}
    \caption{Boresight $75^\circ$.}
    \label{fig:pd_75}
  \end{subfigure}
  \caption{Measured path loss versus TX to RX distance for the wall and RF
           absorber (WA) configuration at boresight angles (a)~$45^\circ$,
           (b)~$60^\circ$, and (c)~$75^\circ$, referenced to
           $\rho_{D\,0} = 1.02$\,m, with the fitted distance exponent
           $n_\mathrm{exp}$ on each panel.}
  \label{fig:pathloss_dist}
  \vspace{-0.25cm}
\end{figure*}
To confirm that the peaks are due to lateral waves propagating along the far face of the drywall, we flushed the RF absorber against the far face, and as a result, the SNR peaks were suppressed as shown in the WCA heatmap (Fig \ref{fig:3d_wca}). The peaks at $57^\circ$ and $69^\circ$ appear due to multipath caused by higher-order specular reflections. The beam reflects from the front-face of the drywall and then from the room walls behind the transceivers or from the surrounding environment before reaching RX. The RF-absorber-lined enclosures do not fully suppress sidelobe coupling at TX-RX separations below $2$\,m. This accounts for the $57^\circ$ and $69^\circ$ returns. These incident rays do not pass through the absorber on the TX side. Instead, RX receives them through its side and back lobes. The front face reflections are suppressed in AO configuration, where the front wall is absent (Fig.~\ref{fig:3d_heatmaps}). The AO configuration also rules out the direct wave and beam overlap between the front wall and the RF absorber behind it. 
 
\paragraph{\textbf{Frequency Scaling}}
We evaluate the path loss by varying the frequency from $60$\,GHz to $69$\,GHz while the separation between TX and RX remains fixed at $1.5$m. Substituting values from Table~\ref{tab:eval_params} into~\eqref{eq:pl_freq} yields
\begin{equation}
  \begin{aligned}
  \Delta PL_{LW}(f) ={}& 40\log_{10}\!\left(\frac{f}{f_0}\right)
    + 5.4\left(\frac{f}{f_0} - 1\right) \\
  &+ 4.343\,\big(K_a(f) - K_a(f_0)\big)\,d_{LW}~\text{dB},
\end{aligned}
  \label{eq:pl_freq_eval}
\end{equation}
where the drywall absorption term, $4.343\,(2t/\cos\theta_k)\,K_w(f_0) = 5.4$\,dB, and the last term is the change in air absorption over the interface leg, $d_{LW} \approx \rho_D = 1.8$\,m, with $K_a(f)$ from the ITU-R~P.676 oxygen model~\cite{itu676attenuation} relative to $K_a(f_0) = 1.7\times10^{-3}$\,Np/m. 

For a reference frequency of $f_0 = 58$\,GHz, the results are shown in Table~\ref{tab:freq_sweep}. The air term
stays within $\pm0.01$\,dB. A least-squares fit of these values to $10\log_{10} f$ yields an effective frequency exponent, $n_f \approx 5.3$, where $4.0$ corresponds to the two spreading terms and $1.3$ to the drywall absorption growth. The air-absorption term reduces the fitted value by only $0.02$. This is well above the $n_f = 2$ expected for a purely in-air return, such as the front-face specular reflection. 
\begin{table}[t]
  \centering
  \caption{Predicted relative path loss over the frequency and distance sweeps.}
  \label{tab:freq_sweep}\label{tab:dist_sweep}
  \begin{tabular}{llllll}
    \hline
    Sweep & Point & Spreading & Drywall & Air & $\Delta PL_{LW}$ \\
          &       & (dB)      & (dB)    & (dB) & (dB) \\
    \hline
    \multirow{5}{*}{$f$ (GHz)}
      & $60$ & $0.59$ & $0.19$ & +$0.01$ & $0.78$ \\
      & $62$ & $1.16$ & $0.37$ & $0.00$ & $1.53$ \\
      & $64$ & $1.71$ & $0.56$ & $0.00$ & $2.27$ \\
      & $66$ & $2.24$ & $0.74$ & -$0.01$ & $2.98$ \\
      & $69$ & $3.02$ & $1.02$ & -$0.01$ & $4.03$ \\
    \hline
    \multirow{4}{*}{$\rho_D$ (m)}
      & $1.02$ & $0.00$ & $-$ & $0.00$ & $0.00$ \\
      & $1.24$ & $1.70$ & $-$ & $0.00$ & $1.70$ \\
      & $1.50$ & $3.35$ & $-$ & +$0.01$ & $3.36$ \\
      & $1.83$ & $5.08$ & $-$ & +$0.01$ & $5.09$ \\
    \hline
  \end{tabular}
\end{table}

In Fig.~\ref {fig:pathloss_freq}, we report the measured path loss over the sweep for the WA configuration at three boresight angles, $45^\circ$, $60^\circ$, and $75^\circ$.  At the boresight angle of  $45^\circ$, we measure $n_\mathrm{exp} = 4.11$, which is above the $n_f = 2$ in-air return, validating the model. At the wider boresight angles of $60^\circ$ and $75^\circ$, the path-loss exponent falls to $1.54$ and $1.83$, respectively, which are close to the free-space value of $2$. At these boresights, TX is oriented more directly toward RX, so the main lobe carries a strong direct and reflected signal over a short TX-–RX path. These components, together with the lateral wave, keep the path loss low, which is why the exponent sits near the free-space value. We can see this in the blocked-lateral-wave case (Fig.~\ref{fig:3d_wca}), where the baseline SNR stays close to that of the direct/reflected signal and the lateral wave enhances the SNR by $1.25$ and $1.54$\,dB.

\paragraph{\textbf{Distance Scaling}}
We experimentally evaluate the path loss by varying the distance between TX and RX from $1.02$m to $1.83$m while the frequency remains fixed at $60$GHz. At a fixed frequency, the slab crossings, the Fresnel coupling, and the constant
offset do not vary with separation, so the range dependence comes only from the
lateral wave's interface leg, and its spreading exponent $n_{LW}$ sets how fast the path loss grows with $\rho_D$. Therefore, we only use the spherical segment exponent, $n_{LW} = 2$, for this geometry. On the other hand, $n_{LW} = 4$ head-wave asymptote in \cite{king2012lateral} is valid for a strong wavenumber contrast across the interface. Since the air-to-drywall refractive index contrast is only $n_w = 1.88$, this assumption does not hold in our case (Section~\ref{sec:channel_model}). Thus, we retain the free-space-like exponent. Accordingly, $\Delta PL_{LW}$ is calculated as,
\begin{equation}
  \Delta PL_{LW}(\rho_D) = 20\log_{10}\!\left(\frac{\rho_D}{\rho_{D\,0}}\right)
    + 4.343\,K_a(f)\,(\rho_D - \rho_{D\,0}),
  \label{eq:pl_dist_eval}
\end{equation}
which leads to a $20$\,dB per decade spreading slope plus a term linear in range, where
$K_a(f) \approx 2.6\times10^{-3}$\,Np/m at $60$\,GHz~\cite{itu676attenuation}. 

In Table~\ref{tab:dist_sweep}, we list the predicted path loss, with $\rho_{D\,0} = 1.02$\,m. A non-zereo leakage coefficient, $\alpha_L$, would add the linear term $4.343\,\alpha_L\,(\rho_D - \rho_{D\,0})$ in addition to the air absorption term.
In Fig.~\ref{fig:pathloss_dist}, we show the measured path loss against log-scale distances at $1.02$, $1.24$, $1.50$ and $1.83$\,m for $45^\circ$, $60^\circ$, and $75^\circ$ boresights, with reference distance of $1.02$\,m. Over these distances, the $45^\circ$ boresight exhibits a path-loss exponent of 1.19, which is below 2, indicating slower attenuation than free-space propagation. Similarly, at boresight angles of $60^\circ$ and $75^\circ$, the path-loss exponents are $0.30$ and $0.83$, respectively. At short separations, the received field consists of the direct wave together with specular reflections and the lateral-wave component, resulting in a low path-loss exponent. This effect becomes more pronounced at higher boresights, where the transmitter beam is directed more toward the receiver, increasing the contribution of the direct wave and further reducing the distance-dependent attenuation.

\textbf{Summary.} The results are consistent with the lateral wave model in Section~\ref{sec:channel_model}. We observe a boresight-stable SNR peak at a grazing angle of $7^\circ$--$9^\circ$ (incidence angle of $81^\circ$--$83^\circ$). At a boresight angle of $45^\circ$, the path-loss exponent is $n_\mathrm{exp} = 4.11$.  This is clearly above $2$, and is qualitatively consistent with the theoretical value of $5.3$. The measured slope carries the two spreading terms of a two-medium path. The distance sweep is consistent with the free-space-like exponent, $n_{LW} = 2$. In both path-loss evaluations, the measured exponents fall below the theoretical values. The gap is largest at the wider boresights, where a strong reflected component over the short range adds to the lateral wave. These deviations are consistent with the first-order model. 

In this work, we demonstrate the effects of drywall, which sits at the low-contrast, low-loss end, whereas concrete, brick, and glass carry higher permittivity and loss, and the stronger contrast would raise the crossing loss and pull the exponent toward $n=4$. In future work, we aim to build on these foundational results by testing additional materials and finishes, as well as additional scenarios such as characterizing corners and layered partition walls. 

\section{Conclusion}
\label{sec:conclusion}

We illustrate a new way to leverage walls for mmWave communication. We experimentally demonstrate the existence of lateral waves through various configurations that suppress direct or reflected waves. Further, we model the frequency and distance path loss of lateral waves and validate the model against experimental results. These results establish the use of walls for mmWave communication as a surface that guides RF energy along it as a lateral wave. This matters most for mmWave communications in indoor settings, where narrow, highly directional beams are often obstructed, and the terminal sits indoors behind a wall. The empirical characterization of mmWave lateral waves lays the foundation for surface-guided links, and motivates studying the behavior of these waves across materials, frequencies, and surface geometries.


\section{Acknowledgement}
\label{sec:acknowledgement}

This work is supported in part by US National Science Foundation (NSF) ECCS-2030272, CNS-2212050 grants.

\vspace{-0.1cm}
\bibliographystyle{IEEEtran}
\bibliography{ref}

\end{document}